\documentclass[a4paper,fleqn]{cas-dc}
\usepackage[numbers,compress]{natbib}

\usepackage{lineno,hyperref,amsmath,amssymb,bm,tikz,multirow}% bold math
\usepackage{subfig}

\def\tsc#1{\csdef{#1}{\textsc{\lowercase{#1}}\xspace}}
\tsc{WGM}
\tsc{QE}
\tsc{EP}
\tsc{PMS}
\tsc{BEC}
\tsc{DE}
\begin{document}
\let\WriteBookmarks\relax
\def\floatpagepagefraction{1}
\def\textpagefraction{.001}

\shorttitle{Real-time generative design of diverse, "truly" optimized structures with controllable structural complexities}

\shortauthors{Z Du et~al.}

\title [mode = title]{Real-time generative design of diverse, "truly" optimized structures with controllable structural complexities}                      

\address[1]{State Key Laboratory of Structural Analysis, Optimization and CAE Software for Industrial Equipment, Department of Engineering Mechanics, Dalian University of Technology, Dalian, 116023, China}

\address[2]{Ningbo Institute of Dalian University of Technology, Ningbo, 315016, China}

\address[3]{Beijing Institute of Spacecraft System Engineering, China Academy of Space Technology, Beijing, 100094, China}

\address[4]{Beijing Institute of Astronautical Systems Engineering, Beijing, 100076, China}

\author[1,2]{Zongliang Du}[orcid=0000-0002-5924-438X]
\cormark[1]
\ead{zldu@dlut.edu.cn}
\cortext[cor1]{Corresponding authors}

\credit{Conceptualization, methodology, supervision, formal analysis, writing-original draft, writing-review and editing, founding acquisition}

\author[1]{Xinyu Ma}
\credit{Methodology, Investigation, validation, visualization, writing-original draft, writing-review and editing}

\author[1]{Wenyu Hao}
\credit{Investigation, validation, visualization, writing-review and editing}

\author[1]{Yuan Liang}
\credit{Investigation, validation, and writing – original draft}

\author[3]{Xiaoyu Zhang}
\credit{Validation, writing-review and editing}

\author[4]{Hongzhi Luo}
\credit{Validation, writing-review and editing}

\author[1,2]{Xu Guo}
\cormark[1]
\ead{guoxu@dlut.edu.cn}
\credit{Conceptualization, methodology, supervision, writing-review and editing, founding acquisition}

\begin{abstract}
Compared with traditional design methods, generative design significantly attracts engineers in various disciplines. In this work, how to achieve the real-time generative design of optimized structures with various diversities and controllable structural complexities is investigated. To this end, a modified Moving Morphable Component (MMC) method together with novel strategies are adopted to generate high-quality dataset. The complexity level of optimized structures is categorized by the topological invariant. By improving the cost function, the WGAN is trained to produce optimized designs with the input of loading position and complexity level in real time. It is found that, diverse designs with a clear load transmission path and crisp boundary, even not requiring further optimization and different from any reference in the dataset, can be generated by the proposed model. This method holds great potential for future applications of machine learning enhanced intelligent design.     
\end{abstract}

% Keywords

\maketitle

\section{Introduction}
\begin{figure*}
	\centering
		\includegraphics[scale=.5]{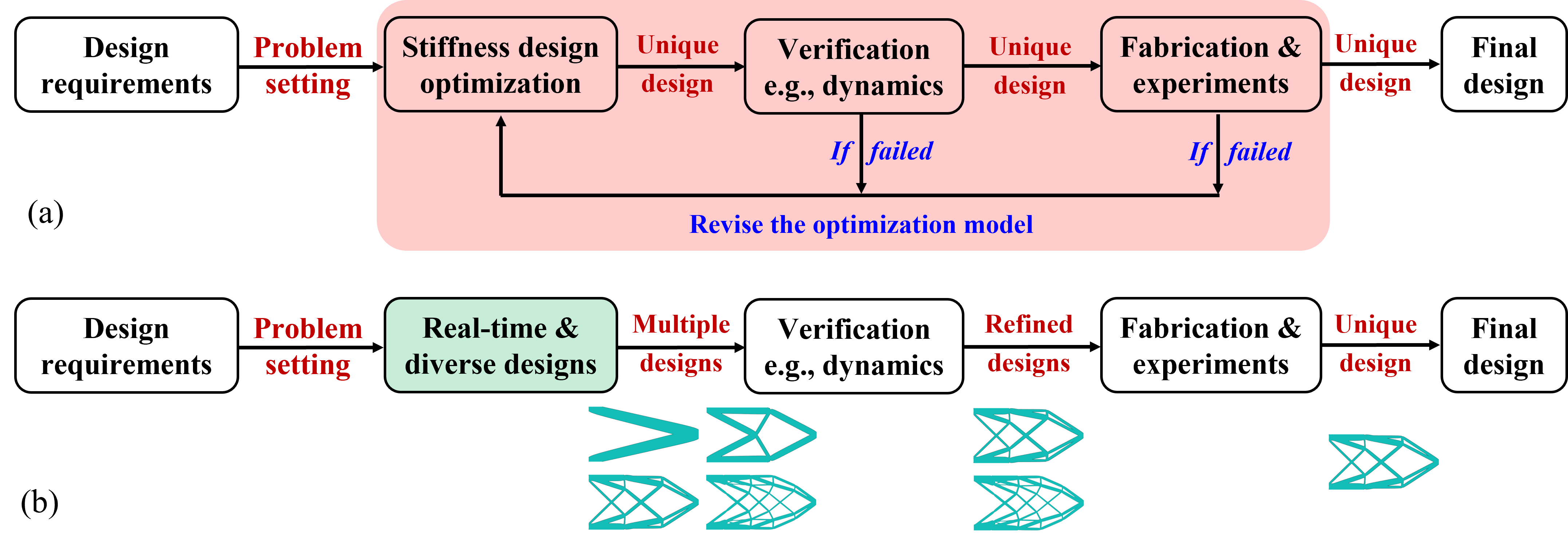}
	\caption{(a) Classic design process of equipment with the help of structural topology optimization; (b) a game-changing design process of equipment with the help of generative design.}
	\label{FIG:1}
\end{figure*}

Structural topology optimization has been well-applied in the innovative design of industrial equipment recently \citep{sigmund2013topology,zhu2016topology}. Nevertheless, it is generally very difficult to take all the factors into account in the topology optimization design process. In practice, as illustrated by Fig. \ref{FIG:1}(a), maximum stiffness or minimum weight design optimization is always solved firstly to produce an conceptual design, and then verification of other performance (e.g., dynamics), fabrication and experiments are evaluated in turns to induce the final design. Since only a unique solution is available in topology optimization method, the highlighted process in Fig. \ref{FIG:1}(a) could be tedious and very inefficient. If diverse optimized designs can be generated in real-time as show by Fig. \ref{FIG:1}(b), the inefficient highlighted process in Fig. \ref{FIG:1}(a) could be fundamentally improved. Obviously, the cornerstone of such game-changing design process is to obtain multiple optimized designs with sufficient diversity in real time, or in other words, to enhance the design process with generative design algorithms \citep{mcknight2017generative}, which aims to automate the generation and evaluation of multiple potential solutions in the design space to identify the best solution. 

With the advances of artificial intelligence, deep learning has been incorporated with topology optimization to accelerate the design process \citep{woldseth2022use}. Based on the moving morphable components (MMC) method \citep{guo2014doing}, Lei et al. investigated to achieve real-time structural topology optimization once the objective/constraint functions and boundary conditions are specified \citep{lei2019machine}. And that work was enhanced with the help of Convolutional Neural Network (CNN) to offer accurate results with a clear structure boundary \citep{geng2023real}. Besides, with the input of the initial displacement field, strain field and volume fraction, a deep CNN was proposed for solving the minimum compliance design problem in negligible time \citep{wang2022deep}. The study in \citep{cang2019one} introduced a theory-driven mechanism for learning a neural network model to establish the mapping between the design variables and the external boundary conditions. Besides establishing an end-to-end relationship between the prescribed optimization parameters and the optimized designs, a problem independent machine learning (PIML) technique, which applies to all design domain and load/boundary conditions, was proposed to reduce the computational time associated with finite element analysis and was able to enhance the solution efficiency of large scale topology optimization problems by orders \citep{huang2022problem,huang2023problem}. 

To obtain diverse designs through structural topology optimization, Zhou et al. proposed alternative formulations for diverse competitive designs with a constraint on performance loss compared to the global optimum \citep{zhou2016balancing}. Later on, graphic diversity measures \citep{wang2018diverse} and a diversity metric \citep{li2021diversity} were incorporated into the optimization formulation to produce multiple designs with diversity. Differently, not modifying the topology optimization formulation, sensitivity dependent strategies (e.g., penalizing certain solid elements of the initial optimized design), sensitivity independent strategies (e.g., modification of the initial design) and stochastic approaches were evaluated to generate diverse and competitive designs \citep{yang2019simple,he2020stochastic,kwok2022improving}. A related topic is controlling the structural complexity in topology optimization process \citep{asadpoure2015incorporating,zuo2022explicit,liu2018parameterized,he2022thinning,zhang2017structural,liang2022explicit}. 

Many researchers also combined the deep-learning based generative model with topology optimization. Oh et al. used topology optimization to create training data, and output designs from Generative Adversarial Networks (GAN) were then used as the reference design for further topology optimization of 2D wheels \citep{oh2019deep}. And such work was extended to generative design of 3D wheels \citep{yoo2021integrating}. Nie et al. later suggested to input the stress and strain energy density fields from the initial material distributions to improve the effectiveness of the generator of a conditional generative adversarial network (CGAN) \citep{nie2021topologygan}. Yu et al. employed CNN to generate low-resolution structures, which are then refined using CGAN to produce high-resolution near-optimal configurations \citep{yu2019deep}. In addition, by inputting domain dimensions, support and loading conditions, and the desired final volume, Kallioras et al. used Deep Belief Networks (DBN) to generate a large number of prototype designs \citep{kallioras2020dzai}. Jang et al. proposed a generative design process based on reinforcement learning (RL) with the objective of maximizing the diversity of topology design \citep{jang2022generative}. The generated design is formulated as a sequence problem to find optimal design parameter combinations based on the given reference design. 

In the above generative design works, mainly using the Solid Isotropic Material with Penalization (SIMP) method \citep{bendsoe2013topology}, the training data are optimized pixel images with gray elements, which inevitably bring challenges to generate black and white designs. Therefore, subsequent optimization is always required to improve the quality of the generated samples. Furthermore, although various topologies were able to provided by the GAN models, the diversity of the outputs was unable to be quantified by some measure, and it is relatively difficult to supply optimized designs with different complexities as illustrated by Fig. \ref{FIG:1}(b). To tackle these challenges, this study proposes a modified MMC method and three strategies controlling structural complexity to produce high quality training data (e.g., optimized designs with a crisp boundary and wide diversity) first. By calculating the topological invariant (i.e., the genus), the optimized designs are classified into different complexity levels. Using the Wasserstein Generative Adversarial Network (WGAN) with an improved loss function, optimized designs with various diversities and structural complexities can be generated in real time, by inputting random noises with the embedding of the boundary condition and structural complexity level. 

The remaining sections of the paper are organized as follows. Section 2 discusses the adopted topology optimization method, novel strategies and complexity measure for generating a high-quality dataset. Section 3 presents the architecture and loss function of our WGAN model and some training details. Section 4 presents the numerical results. Finally, Section 5 summarizes the conclusions and discusses future work.

\section{Generation of high-quality dataset by the modified MMC topology optimization method}
To train a well-performed generative model of optimized structures, a high quality training data, e.g., optimized designs with crisp boundary and wide diversity, is necessary. To achieve this, we first introduce an explicit topology optimization method and three strategies to produce optimized designs with clear load transmission paths and different similarities. With a complexity index measured by the genus, high-quality samples are prepared and labelled by the loading positions and complexity level.  

\subsection{Introduction to the modified MMC method}
\subsubsection{Explicit structural description}
\begin{figure*}
	\centering
		\includegraphics[scale=0.6]{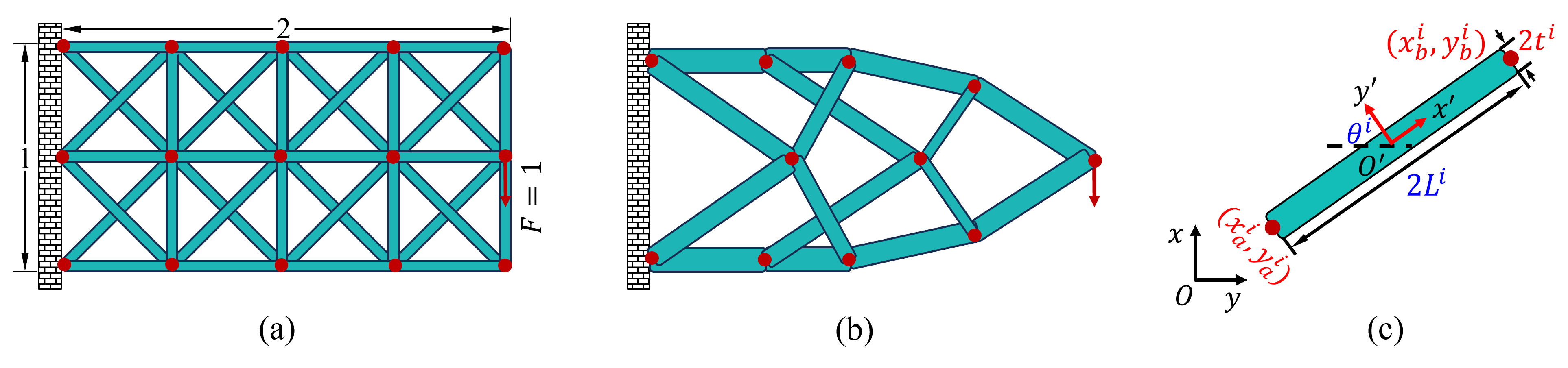}
	\caption{(a) Initial design of the node-driven based MMC method (red points - driven nodes); (b) the corresponding optimized design (the components with 0 thickness is eliminated); (c) a component is totally determined by five geometry parameters $\left( x_a^i, y_a^i, x_b^i, y_b^i, t^i \right)^\top$. The dimensionless size of design domain is $2 \times 1$, the amplitude of the external load is $1$, and the elstic constants of the base material is $E=1, \nu=0.3$, respectively.}
	\label{FIG:2}
\end{figure*}

The building blocks in the MMC method is a set of moving morphable components described by explicit geometry parameters  \citep{guo2014doing,zhang2016new,guo2016explicit}. In particular, the solid region occupied by the $i$-th component (i.e., $\Omega^i$) is identified by its topology description function (TDF) $\phi^i$ as following:
\begin{equation}
\begin{cases}
\phi^i(\boldsymbol{x})>0, & \text { if } \boldsymbol{x} \in \Omega^i \cap \mathrm{D} \\ 
\phi^i(\boldsymbol{x})=0, & \text { if } \boldsymbol{x} \in \partial \Omega^i \cap \mathrm{D} \\ 
\phi^i(\boldsymbol{x})<0, & \text { if } \boldsymbol{x} \in \mathrm{D} \backslash\left(\Omega^i \cup \partial \Omega^i\right)
\end{cases}
\label{eq1}
\end{equation}
where $\mathrm D$ is the design domain. 

To produce optimized designs with a clear load transmission path, a node-driven based MMC method (or the adaptive ground structure approach) is used \citep{jiang2022unified}. In this method, as illustrated by Fig. \ref{FIG:2}, some nodes are distributed in the design domain. By connecting the nodes, straight components with uniform widths are assembled as a structure. The connection relation of nodes keeps the same, and only the locations of nodes and the widths of components are iteratively updated in optimization process. In particular, for the $i$-th component in Fig. \ref{FIG:2}(c), its half-length $L^i$, center coordinates $(x_0^i,y_0^i)$, sine and cosine values of its inclined angle $\theta^i$ are determined by its end points as 
\begin{equation}
\begin{cases}
L^i=\frac{1}{2}\left(\left(x_a^i-x_b^i\right)^2+\left(y_a^i-y_b^i\right)^2\right)^{\frac{1}{2}} \\
x_0^i=\frac{x_a^i+x_b^i}{2} \quad y_0^i=\frac{y_a^i+y_b^i}{2} \\
\sin \theta^i=\frac{y_a^i-y_b^i}{2 L^i} \quad \cos \theta^i=\frac{x_a^i-x_b^i}{2 L^i} 
\end{cases}
\label{eq2}
\end{equation}

By placing a local Cartesian coordinate system $O'x'y'$ at the center of the $i$-th component, the corresponding TDF can be constructed as follows \citep{du2022efficient}:
\begin{equation}
\phi^i =1 -\left(\left(\frac{x^{\prime}}{L^i}\right)^p+\left(\frac{y^{\prime}}{t^i}\right)^p\right)^{\frac{1}{p}}
\label{eq3}
\end{equation}
with
\begin{equation}
\begin{cases}
x^{\prime}=\cos \theta^i \left(x-x_0^i\right)+\sin \theta^i\left(y-y_0^i\right) \\
y^{\prime}=-\sin \theta^i \left(x-x_0^i\right)+\cos \theta^i\left(y-y_0^i\right)
\end{cases}
\label{eq4}
\end{equation}
where $t^i$ denote the half-width, and $p$ is a relatively large even number (e.g., $p=6$). In this manner, the $i$-th component is explicitly described by five geometric parameters, i.e., ${\bm d}_i=\left( x_a^i, y_a^i, x_b^i, y_b^i, t^i \right)^\top$.

With all TDFs of components obtained, the TDF of the entire structure can be expressed by the K-S function to approximate the max operation \citep{liu2018efficient}:
\begin{equation}
\phi^{\mathrm{s}} = \max \left(\phi^1,...,\phi^N \right) 
\approx\left(\ln \left(\sum_{i=1}^N \exp \left(\lambda \phi^i\right)\right)\right) / \lambda
\label{eq5}
\end{equation}
where $N$ is the number of MMCs and $\lambda$ is a relatively large even number, e.g., $\lambda=80$. 

Similar as Eq. (\ref{eq1}), the structural region can be identified as the region $\Omega=\left\{{\bm x} \mid {\bm x} \in \mathrm{D}, \phi^{\mathrm{s}}(\boldsymbol{x})>0\right\}$. Supposing there are totally $M$ driven nodes and $N$ components in the design domain, the design variable can be expressed as $\boldsymbol{d} = \left({\bm d}_1^\top,...,{\bm d}_n^\top \right)^\top = \left({\bm d}_x^{\top}, {\bm d}_y^{\top}, {\bm d}_t^{\top} \right)^\top$, where ${\bm d}_x = \left(x_1,...,x_M \right)^{\top}$, ${\bm d}_y = \left(y_1,...,y_M \right)^{\top}$, and ${\bm d}_t = \left(t_1,...,t_N \right)^{\top}$. The total number of design variables is $N + 2M$.

It is worth noting that, the adopted topology optimization approach offers three key benefits: (1) during the optimization process, structural connection relationships are maintained, and guarantees that the optimization results have a clear load transmission paths and crisp boundary; (2) by controlling the number of nodes and components in the initial designs, the complexity of optimized designs can be conveniently controlled as shown in subsection 2.2; (3) the modified MMC method is well-suited for integration with CAD system, making its advantages even more apparent during subsequent design and application process.

\subsubsection{Mathematical formulation}
For minimum compliance design problems, the mathematical formulation with finite element discretization is expressed as:
\begin{equation}
\begin{aligned}
\text {find } \quad & \boldsymbol{d} \in \mathcal{U}^{\boldsymbol{d}} \\
\min \, \quad & f=\boldsymbol{F}^{\top} \boldsymbol{U} \\
\text { s.t. } \, \quad & \mathbf{K} \boldsymbol{U}=\boldsymbol{F} \\
& g=V /|\mathrm{D}| - \bar{v} \leq 0
\end{aligned}
\label{eq6}
\end{equation}
where ${\mathbf{K}},\bm{F}$ and $\bm{U}$ are the global stiffness matrix, nodal force vector, and nodal displacement vector; $V, |\mathrm{D}|, \bar{v}$ are the volume of the optimized structure, the volume of the design domain, the upper bound of allowable volume fraction and $\mathcal{U}^{\boldsymbol{d}} $ denotes the admissible set of design variable, respectively.

In the finite element analysis with Eulerian mesh, the ersatz material model is used \citep{zhang2016new, du2022efficient}. Specifically, the global stiffness matrix is assembled using the elemental stiffness matrix of $e$-th element as
\begin{equation}
    {\bf k}_e = \rho_e {\bf k}_e^0 = \frac{{\bf k}_e^0}{4} \sum_{i=1}^{4} H_\epsilon^\alpha \left(\phi_{e,i}^{\rm s} \right)
    \label{eq7}
\end{equation}
where $\rho_e$ and ${\bf k}_e^0$ are the density and the elemental stiffness matrix of base material of the $e$-th element, respectively. The symbol $\phi_{e,i}^{\rm s}$ is TDF value of the entire structure on the $i$-th node of $e$-th element. The regularized Heaviside function $H_\epsilon^\alpha$ is expressed as
\begin{equation}
    H_\epsilon^\alpha(x) = 
    \begin{cases}
    1, & \text {if } x > \epsilon \\ 
    \frac{3\left(1-\alpha\right)}{4}\left(\frac{x}{\epsilon}-\frac{x^3}{3\epsilon^3}\right)+\frac{1+\alpha}{2}, & \mathrm{if} -\epsilon\leq{x}\leq\epsilon \\ 
    \alpha, & \text {otherwise }
    \end{cases}
    \label{eq8}
\end{equation}
In Eq. (\ref{eq8}), $\epsilon=0.1$ is the width of regularized region and
$\alpha=10^{-3}$ is a small positive number, respectively.

\subsubsection{Sensitivity analysis}
For the concerned objective and constraint functions in Eq. (\ref{eq6}), the corresponding sensitivities with respect to a specific design variable $d_i$ ($i=1,...,N+2M$, the $i$-th parameter of the design variable vector related to totally $P_i$ components in the index set $\mathcal{S}_i$) can be calculated as:
\begin{equation}
\begin{cases}
    \frac{\textstyle \partial f}{\textstyle \partial d_i}   & = \displaystyle \sum_{e = 1}^{Ne} \frac{\partial f}{\partial \rho_e} \frac{\partial \rho_e}{\partial d_i}\\
    & = \displaystyle \sum_{e = 1}^{Ne} \boldsymbol{u}_e^{\top} \mathbf{k}_e^0 \boldsymbol{u}_e \frac{\partial \rho_e}{\partial \phi^{\rm s}} \left(  \sum_{j = 1}^{P_j} \frac{\partial \phi^{\rm s}}{\partial \phi^{\mathcal {S}_{i,j}}} \frac{\partial \phi^{\mathcal {S}_{i,j}}}{\partial d_i}\right)\\
    \frac{\textstyle \partial g}{\textstyle \partial d_i}   & = \displaystyle \sum_{e = 1}^{Ne} \frac{\partial g}{\partial \rho_e} \frac{\partial \rho_e}{\partial d_i}\\
    & = \displaystyle \sum_{e = 1}^{Ne} \frac{V_e}{|\mathrm{D}|} \frac{\partial \rho_e}{\partial \phi^{\rm s}} \left( \sum_{j = 1}^{P_j} \frac{\partial \phi^{\rm s}}{\partial \phi^{\mathcal {S}_{i,j}}} \frac{\partial \phi^{\mathcal {S}_{i,j}}}{\partial d_i}\right)
\end{cases}
\label{eq9}
\end{equation}
where $Ne$, ${\bm u}_e$, and $V_e$ are the total number of finite elements, the nodal displacement vector and the volume of the $e$-th element, respectively. The symbol $\phi^{\mathcal {S}_{i,j}}$ refers to the TDF of the $j$-th component in the index set $\mathcal{S}_i$. 

According to Eqs. (\ref{eq5}) and (\ref{eq7}), we have 
\begin{equation}
    \frac{\partial \rho_e}{\partial \phi^{\rm s}} \frac{\partial \phi^{\rm s}}{\partial \phi^{\mathcal {S}_{i,j}}} = \frac{1}{4} \sum_{k=1}^{4} H_\epsilon^{\alpha \prime} \left(\phi_{e,k}^{\rm s} \right) \frac{\exp \left({\lambda \phi^{\mathcal {S}_{i,j}}_{e,k}} \right)}{\sum_{l=1}^N \exp \left({\lambda \phi^l} \right)}
\label{eq10}
\end{equation}
where $^\prime$ denotes the differential operator. For the conciseness of the main text, the exact expressions of $\partial \phi^{\mathcal {S}_{i,j}} / \partial d_i$ are presented in Appendix \ref{apdx1}.

\begin{table*}[width=2.05\linewidth,cols=7,pos=h]
\caption{Generating diverse optimized designs by Strategy 1: introducing different numbers of driven nodes.}\label{tbl1}
\begin{tabular*}{\tblwidth}{@{} CCCCC@{} }
\toprule
Initial designs & \begin{minipage}[b]{0.37\columnwidth}
    \centering
    \raisebox{-.5\height}{\includegraphics[width=\linewidth]{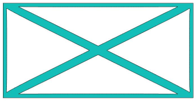}}
\end{minipage} & \begin{minipage}[b]{0.37\columnwidth}
    \centering
    \raisebox{-.5\height}{\includegraphics[width=\linewidth]{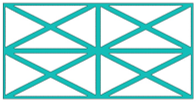}}
\end{minipage} & \begin{minipage}[b]{0.37\columnwidth}
    \centering
    \raisebox{-.5\height}{\includegraphics[width=\linewidth]{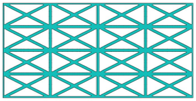}}
\end{minipage} & \begin{minipage}[b]{0.37\columnwidth}
    \centering
    \raisebox{-.5\height}{\includegraphics[width=\linewidth]{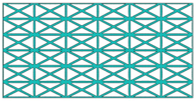}}
\end{minipage}  \\
\\
Optimized designs & \begin{minipage}[b]{0.37\columnwidth}
    \centering
    \raisebox{-.5\height}{\includegraphics[width=\linewidth]{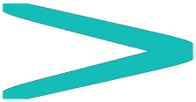}}
\end{minipage} & \begin{minipage}[b]{0.37\columnwidth}
    \centering
    \raisebox{-.5\height}{\includegraphics[width=\linewidth]{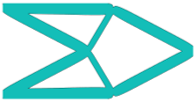}}
\end{minipage} & \begin{minipage}[b]{0.37\columnwidth}
    \centering
    \raisebox{-.5\height}{\includegraphics[width=\linewidth]{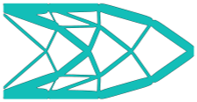}}
\end{minipage} & \begin{minipage}[b]{0.37\columnwidth}
    \centering
    \raisebox{-.5\height}{\includegraphics[width=\linewidth]{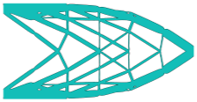}}
\end{minipage}  \\
\\
Compliance values & $110.4$ & $87.0$ & $84.5$ & $83.3$  
\\\bottomrule
\end{tabular*}
\end{table*}

\subsection{Three strategies for generating optimized designs with diversities}
On the one hand, one of the goals of generative design is to provide designers with various reference structures to choose from. On the other hand, datasets composed of diverse structures are beneficial for training, increasing GAN's stability and avoiding mode collapse. To enhance the diversity of the dataset, we introduce three strategies for the solution process of formulation (\ref{eq6}). 
\subsubsection{Strategy 1: Diverse designs by introducing different numbers of driven nodes}
Since new driven nodes and components cannot be introduced in the optimization process, the complexity of an optimized design is determined by the initial number of driven nodes to some extent. In particular, for the cantilever beam example illustrated by Fig. \ref{FIG:2}(a), the maximum iteration number is fixed as 150. Fixed the base cell as the first initial design in Table \ref{tbl1}, arrays from $1 \times 1$ to $7 \times 7$ base cells were set as initial designs to obtain optimized designs. It shows that as the driven nodes and components in the initial designs get more, more complex optimized designs are obtained with a smaller compliance values. Notably, when initial design is too simple, highly inferior optimized design may be obtained (the first column in Table \ref{tbl1}).

\begin{table*}[width=2.05\linewidth,cols=7,pos=h]
\caption{Generating diverse optimized designs by Strategy 2: setting different initial node positions.}\label{tbl2}
\begin{tabular*}{\tblwidth}{@{} CCCCC@{} }
\toprule
Preoptimized iterations & $5$ & $15$ & $25$ & $35$  
\\
\\
Initial designs & \begin{minipage}[b]{0.37\columnwidth}
    \centering
    \raisebox{-.5\height}{\includegraphics[width=\linewidth]{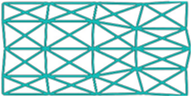}}
\end{minipage} & \begin{minipage}[b]{0.37\columnwidth}
    \centering
    \raisebox{-.5\height}{\includegraphics[width=\linewidth]{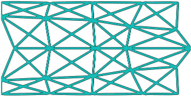}}
\end{minipage} & \begin{minipage}[b]{0.37\columnwidth}
    \centering
    \raisebox{-.5\height}{\includegraphics[width=\linewidth]{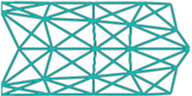}}
\end{minipage} & \begin{minipage}[b]{0.37\columnwidth}
    \centering
    \raisebox{-.5\height}{\includegraphics[width=\linewidth]{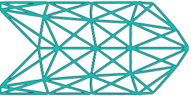}}
\end{minipage}  \\
\\
Optimized designs & \begin{minipage}[b]{0.37\columnwidth}
    \centering
    \raisebox{-.5\height}{\includegraphics[width=\linewidth]{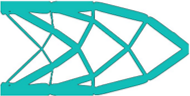}}
\end{minipage} & \begin{minipage}[b]{0.37\columnwidth}
    \centering
    \raisebox{-.5\height}{\includegraphics[width=\linewidth]{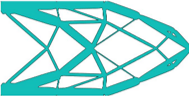}}
\end{minipage} & \begin{minipage}[b]{0.37\columnwidth}
    \centering
    \raisebox{-.5\height}{\includegraphics[width=\linewidth]{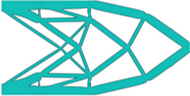}}
\end{minipage} & \begin{minipage}[b]{0.37\columnwidth}
    \centering
    \raisebox{-.5\height}{\includegraphics[width=\linewidth]{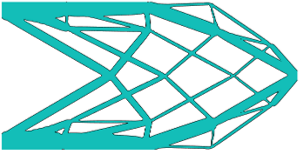}}
\end{minipage}  \\
\\
Compliance values & $84.4$ & $85.5$ & $84.3$ & $84.2$  
\\\bottomrule
\end{tabular*}
\end{table*}

\subsubsection{Strategy 2: Diverse designs by setting different locations of driven nodes}
Since completely random distribution of node positions lacks a certain degree of rationality, a method has been developed to ensure the rationality of the node positions while introducing randomness to the initial layout. First, the initial layouts in Strategy 1 are pre-optimized for random iterations (1 to 40) with the sensitivity values of thickness setting as 0 to keep the width of the components as constant. Then the corresponding designs are set as initial design of Strategy 2, as the Table \ref{tbl2}. Those initial designs are further optimized by 150 iterations with no restriction on sensitivities to obtain optimized designs with close compliance values. Generally, as the pre-optimized iteration gets larger, more components are retrained to obtain optimized designs with a larger complexity. 

\begin{table*}[width=2.05\linewidth,cols=7,pos=h]
\caption{Generating diverse optimized designs by Strategy 3: adding extra components (20 components are added).}\label{tbl3}
\begin{tabular*}{\tblwidth}{@{} CCCCC@{} }
\toprule
Initial designs & \begin{minipage}[b]{0.37\columnwidth}
    \centering
    \raisebox{-.5\height}{\includegraphics[width=\linewidth]{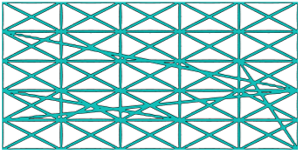}}
\end{minipage} & \begin{minipage}[b]{0.37\columnwidth}
    \centering
    \raisebox{-.5\height}{\includegraphics[width=\linewidth]{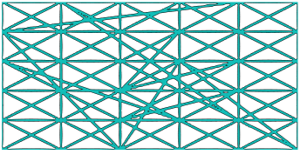}}
\end{minipage} & \begin{minipage}[b]{0.37\columnwidth}
    \centering
    \raisebox{-.5\height}{\includegraphics[width=\linewidth]{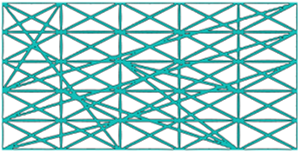}}
\end{minipage} & \begin{minipage}[b]{0.37\columnwidth}
    \centering
    \raisebox{-.5\height}{\includegraphics[width=\linewidth]{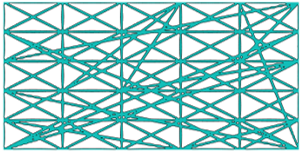}}
\end{minipage}  \\
\\
Optimized designs & \begin{minipage}[b]{0.37\columnwidth}
    \centering
    \raisebox{-.5\height}{\includegraphics[width=\linewidth]{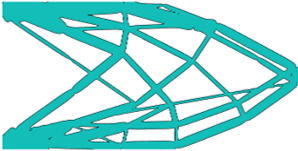}}
\end{minipage} & \begin{minipage}[b]{0.37\columnwidth}
    \centering
    \raisebox{-.5\height}{\includegraphics[width=\linewidth]{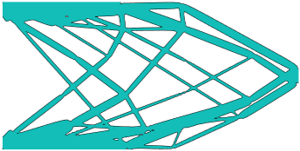}}
\end{minipage} & \begin{minipage}[b]{0.37\columnwidth}
    \centering
    \raisebox{-.5\height}{\includegraphics[width=\linewidth]{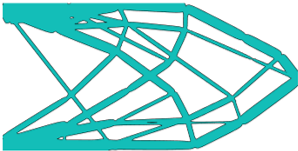}}
\end{minipage} & \begin{minipage}[b]{0.37\columnwidth}
    \centering
    \raisebox{-.5\height}{\includegraphics[width=\linewidth]{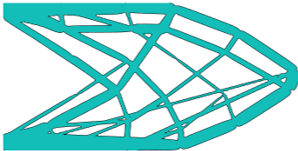}}
\end{minipage}  \\

\\
Compliance values & $84.3$ & $85.5$ & $85.0$ & $84.6$  
\\\bottomrule
\end{tabular*}
\end{table*}

\subsubsection{Strategy 3: Diverse designs by adding extra components}
In addition to modify the initial layout and position of nodes, another way to introduce diversity is by randomly adding new connection relationships (extra components) to the initial layout. On the one hand, the basic layout ensures that the resulting structure complexity is similar. On the other hand, the added connection relationships guide the structure evolving toward diverse results. This approach can also achieve the goal of obtaining diverse structures with a similar complexity. Table \ref{tbl3} shows the initial designs by adding 20 random extra components and the corresponding optimized configurations based on a $5 \times 5$ arrayed base cells. Notably, although randomly adding extra components may break the symmetry of optimized design, nevertheless, the optimized structural compliance is not increased too much. 

\subsection{Measurement of structural complexity based on the topological invariant}
Optimized designs obtained through the above methods includes structures ranging from simple to complex, but how can we quantitatively measure the complexity of the structure and embed the measure into neural networks? Actually, the structural complexity of the optimized designs can be measured by their topology invariants (i.e., Euler and Betti numbers for 2D and 3D cases respectively). To this end, we adopt the algorithm proposed in Liang et al. to efficiently calculate the topological number of optimized designs \citep{liang2022explicit}. 

\begin{figure}
	\centering
		\includegraphics[scale=.35]{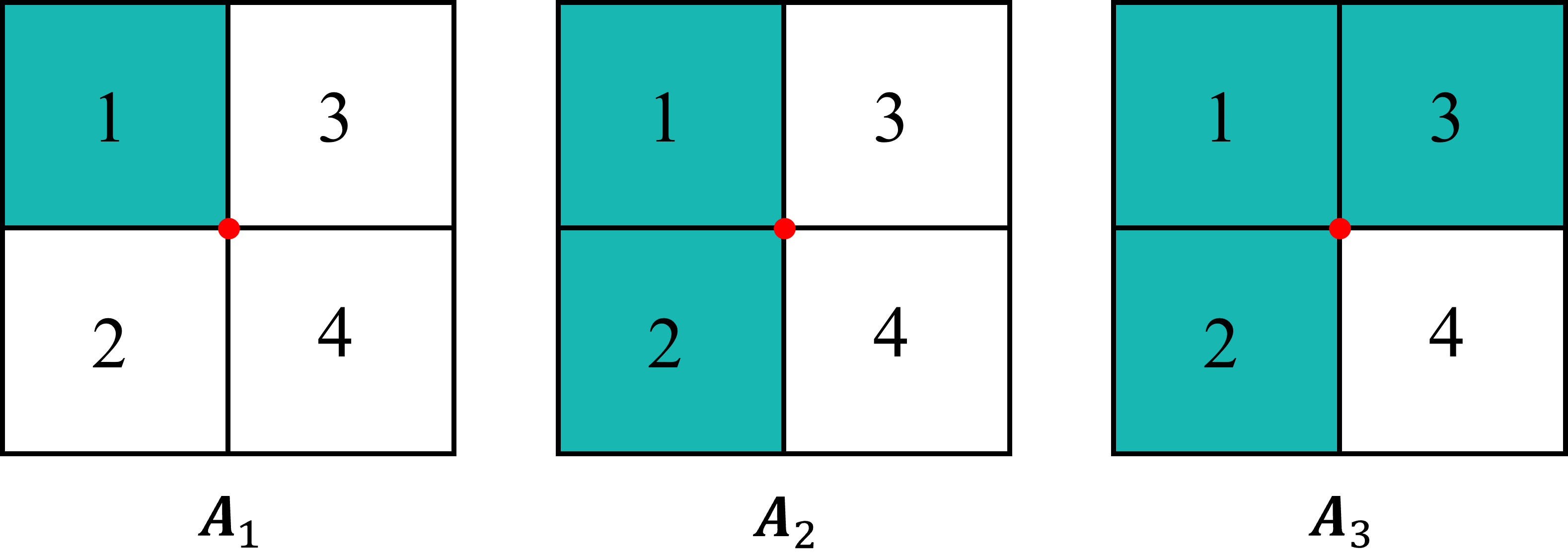}
	\caption{The cases account for $|{\bm A}_1|$, $|{\bm A}_2|$ and 
 $|{\bm A}_3|$ (number of elements with nodal density $1/4$, $2/4$ and $3/4$, respectively).}
	\label{FIG:3}
\end{figure}

Firstly, the optimized structures are converted into a 0-1 pixel-based images based on its density values of each element for FEA. Then, using the programmable Euler–Poincaré formula expressed as
\begin{equation}
E=B_0-B_1=\frac{|{\bm A}_1|-|{\bm A}_3|}{4}
\label{eq11}
\end{equation}
where $E$ denotes the Euler number, $B_0$ and $B_1$ are two Betti numbers, i.e., the number of the independent connected component and the genus (i.e., the number of interior holes), respectively. In implementation, the symbols $|{\bm A}_1|$ and 
 $|{\bm A}_3|$ count the number of elements with nodal density $1/4$ and $3/4$, as shown in the Fig. \ref{FIG:3}. The number of the independent connected component $B_0$ is 1 due to the requirement of the maximum stiffness. The the genus $B_1$ can be calculated directly. More details is refereed to \cite{liang2022explicit}. 

\begin{figure}
	\centering
		\includegraphics[scale=.17]{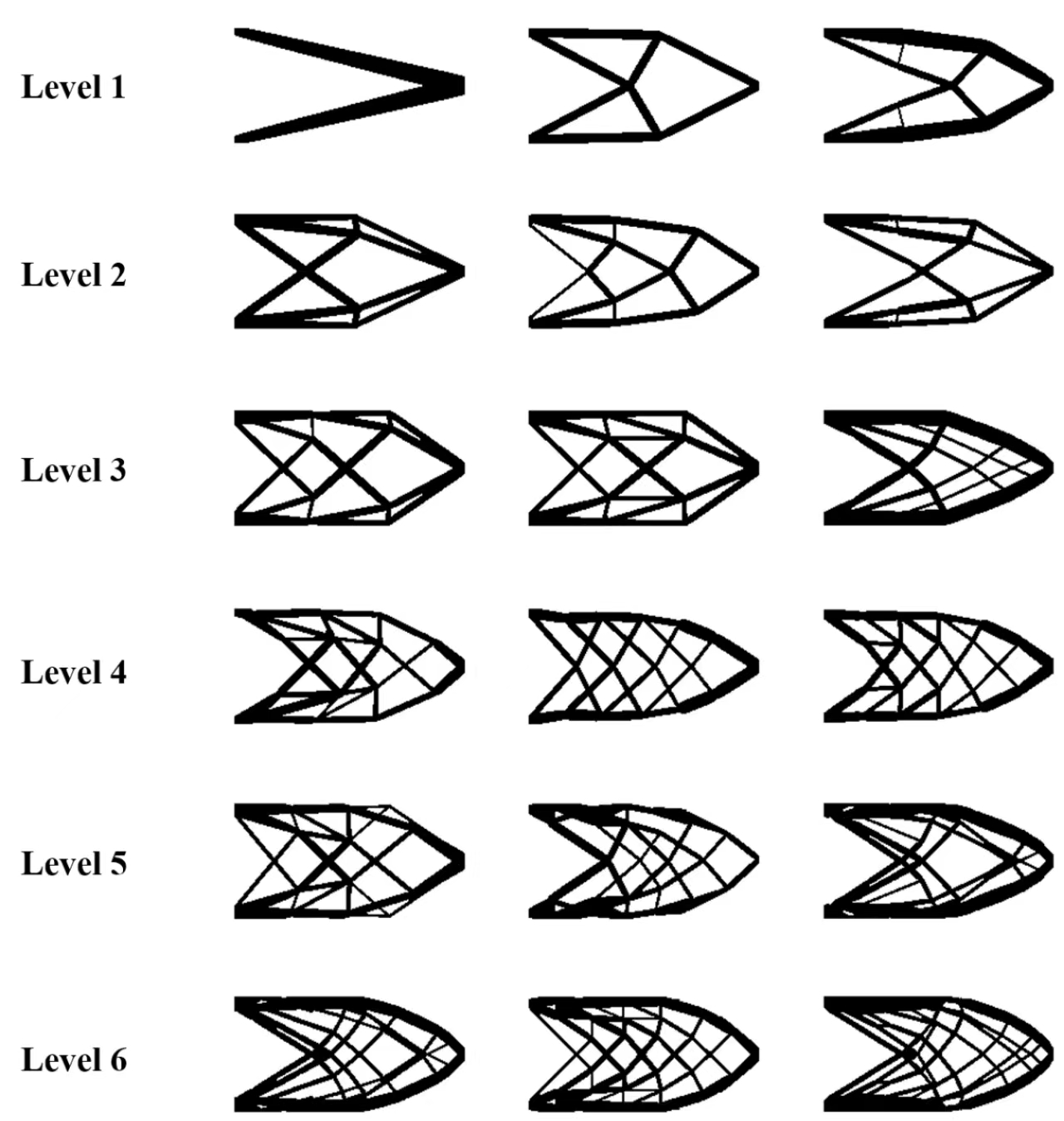}
	\caption{Some typical optimized designs classified by the complexity levels.}
	\label{FIG:4}
\end{figure}

It should be noted that, directly using the genus as the complexity measure could encounter two main difficulties in generative design: first, the genus could be too fine to measure the complexity and similarity of optimized structures. For instance, the genus of the first and the fourth optimized design in the Table \ref{tbl3} is 29 and 34, respectively. Nevertheless, they are quite similar and have close compliance values. Besides, setting the genus as a label generally requires a large dataset to train the neural network model. Therefore, we suggest to classify the complexity levels depend on the genus, e.g., complexity level 1 for genus from 0 to 5, complexity level 2 for genus from 6 to 10, ..., complexity level 6 for genus over 25. Fig. \ref{FIG:4} presents some representative structures classified by the complexity levels. In this manner, diverse structures can be generated with different complexity levels. The effectiveness of this complexity measure will be demonstrated by the numerical examples later.

\subsection{Dataset preparation}

\begin{figure*}[h]
	\centering
		\includegraphics[scale=.2]{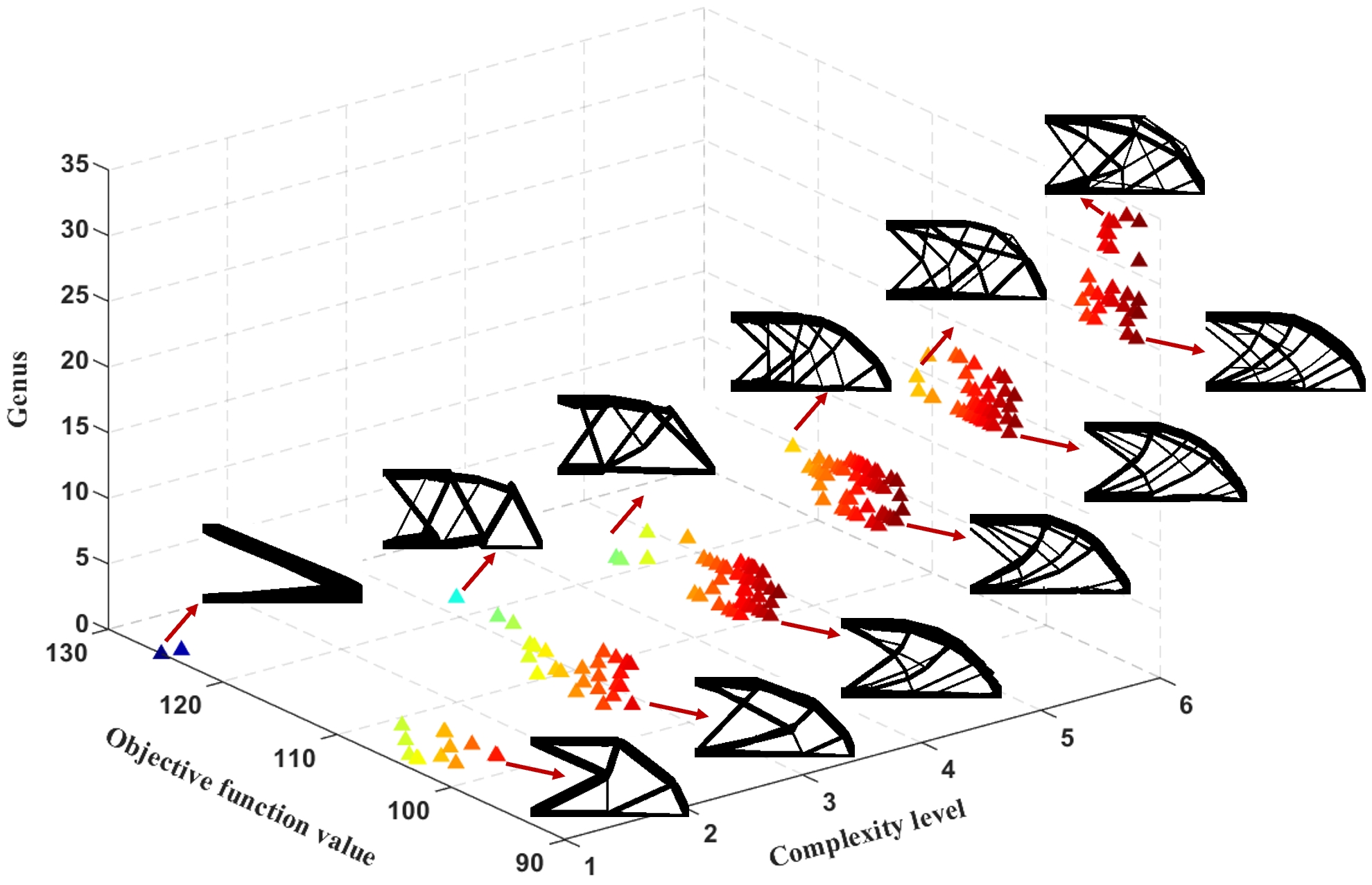}
	\caption{Distribution of the objective function value, complexity level and genus of samples corresponding to loading position 0.}
	\label{FIG:5}
\end{figure*}

\begin{figure*}[h]
	\centering
		\includegraphics[scale=.2]{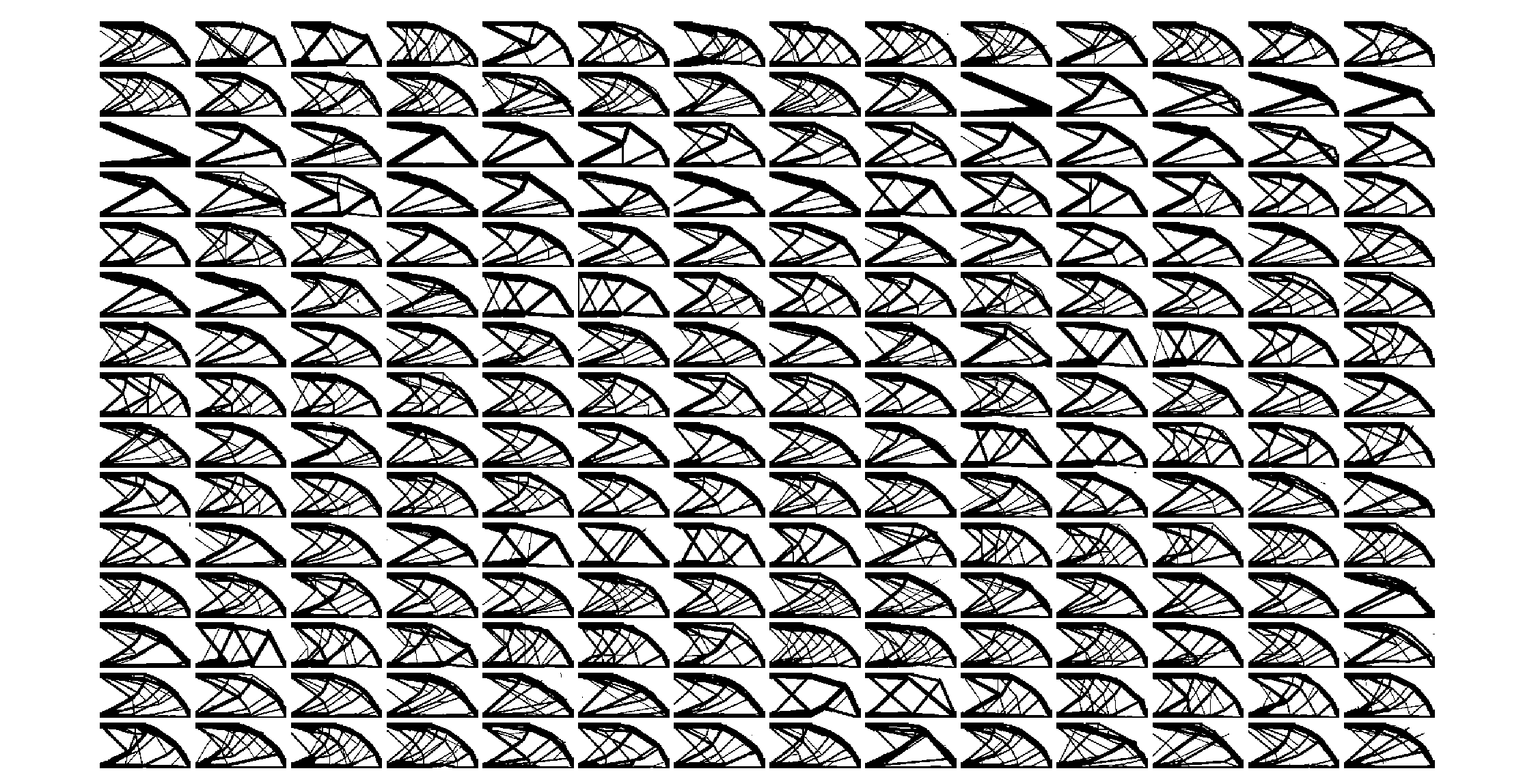}
	\caption{Samples of optimized designs corresponding to location 0.}
	\label{FIG:6}
\end{figure*}

By employing strategies 1, 2, and 3, a diverse dataset of optimized structures is generated. Taking the cantilever beam problem in Fig. \ref{FIG:2}(a) as an example, with a grid size of $200 \times 100$, the dataset includes optimized designs subject to a concentrated load at 101 possible loading position (the grid nodes along the right boundary). Through the aforementioned method and strategies, approximately 190 structures are obtained for each loading position and the dataset consists of a total of 19,015 images. To ensure uniformity in network size for different problem settings and for ease of processing, the optimized designs are adjusted to $200 \times 200$ pixel images.  Using Eq. (\ref{eq11}), the topological number of each structure is calculated, and complexity levels are assigned.

To validate the effectiveness of the dataset, Fig. \ref{FIG:5} presents a scatter diagram for the dataset with loading position 0 (bottom right corner), including structures' objective function value, genus, and complexity level. For each complexity level, it showcases the optimized structures with the best and worst objective functions. It can be observed intuitively that as the genus of structures increases, the complexity of the structures also rises, and the overall performance of the structures gradually improves.

For instance, in the case of structures with a complexity level of 1, the majority of designs' objective function values are around 100. Particularly, the simplest structure, with genus of 0, exhibits an objective function value about 125. For structures with a complexity level of 6, the structural compliance values for most configurations are in the range of 92 to 95. It is noteworthy that the number of structures with a complexity level of 1 is relatively small, which is reasonable. Considering that at such small genus (0 to 5), the possible configurations of optimized structures are limited. Therefore, our data set not only encompasses optimized structures ranging from simple to complex but also illustrates diverse configurations under similar complexity levels. Meanwhile, Fig. \ref{FIG:6} illustrates the entire dataset at loading position 0, revealing that there are no particularly similar configurations in the dataset. It not only encompasses structures ranging from simple to complex but also diverse structures under similar complexity levels. This demonstrate the effectiveness of the proposed method in generating a high quality training dataset for generative model. The related dataset can be found in \href{https://github.com/ZongliangDu/Data-set-for-Real-time-generative-design-of-optimized-structures}{https://github.com/ZongliangDu/Data-set-for-Real-time-generative-design-of-optimized-structures}.

\section{An improved WGAN model for real-time generative design}
In this section, after giving a brief introduction about the WGAN, some modifications on the architecture and loss function are proposed for the adopted WGAN model. 
\begin{figure*}[h]
	\centering
		\includegraphics[scale=1.1]{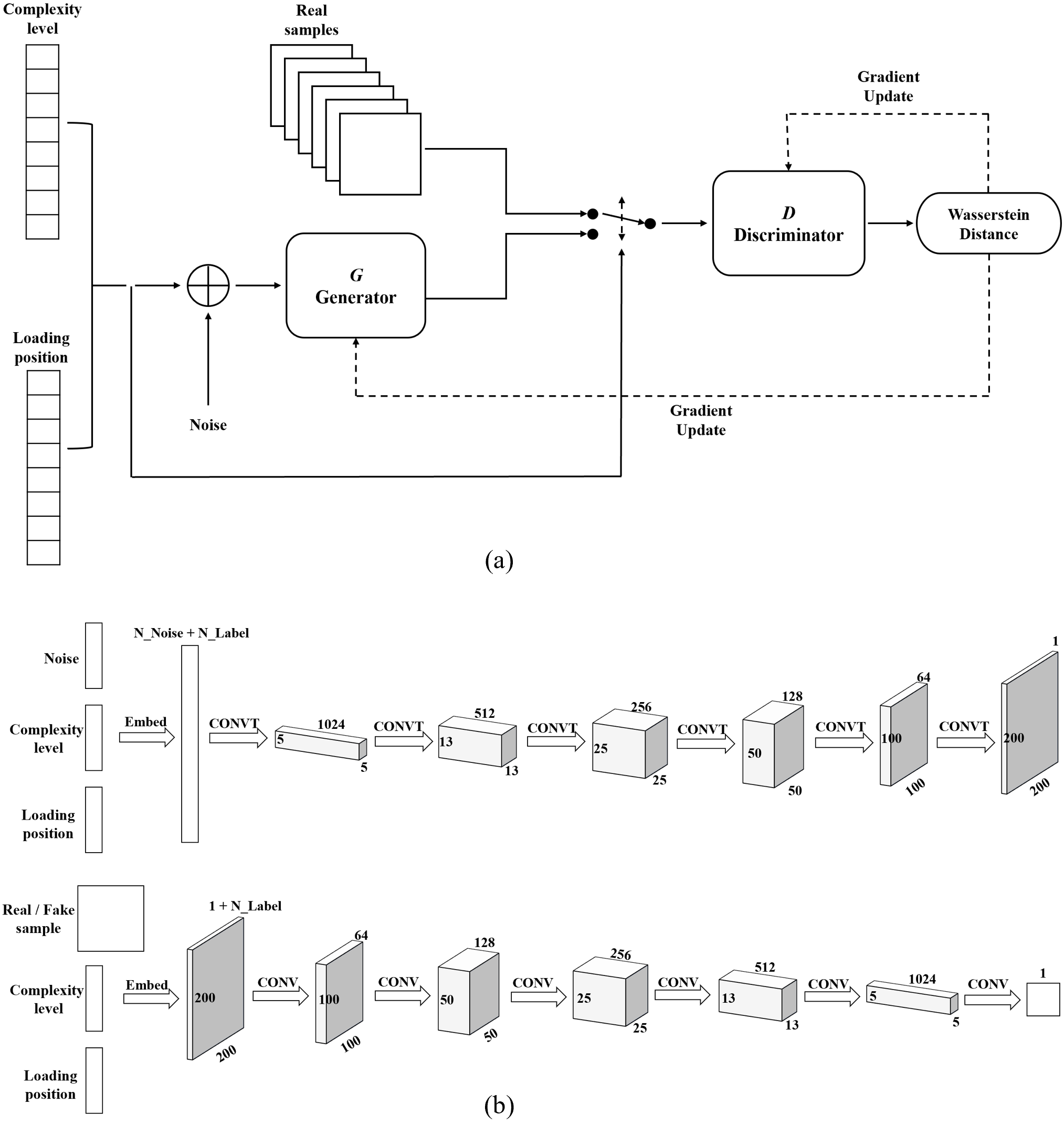}
	\caption{(a) Architecture of the adopted WGAN model; (b) details of the generator and the discriminator.}
	\label{FIG:7}
\end{figure*}

\subsection{Introduction of the WGAN}
GAN is a highly versatile generative model that produces diverse, complex outputs like images, audio, and text. The celebrated GAN model is composed of two components: the generator and the discriminator \citep{goodfellow2020generative}. The generator aims to generate data that resembles the real samples, while the discriminator aims to differentiate between the generated data and real data. The two networks are trained using an adversarial loss function, where the accuracy of the discriminator increases as the generated data becomes more realistic. During the training process, the generator continually strives to produce increasingly realistic data, while the discriminator continually improves its ability to identify whether the data is real or generated.

Wasserstein Generative Adversarial Network (WGAN) is a modification of the standard GAN \citep{arjovsky2017wasserstein}. Different as traditional GANs, which use the binary cross-entropy loss as the loss function, WGANs utilize the Wasserstein distance, also called Earth Mover's Distance (EMD), as the adversarial loss function. This metric measures the distance between two probability distributions, leading to a more robust and stable training process. Additionally, WGANs enforce constraints on the discriminator network's weights, ensuring its Lipschitz continuity, which further enhances the stability of the training. During our research, we discovered that models trained with WGAN exhibit better stability and can produce outputs with a higher quality.

\subsection{Improved WGAN model for real-time generative design with various diversity and structural complexities}
\subsubsection{Architecture of the adopted WGAM model}
The architecture of our WGAN model is depicted in Fig. \ref{FIG:7}(a). The details of the generator $G$ and the discriminator $D$ are shown in Fig. \ref{FIG:7}(b). In general, the networks of $G$ and $D$ can be both convolutional neural networks and the setting of dimensions has duality.

For the generator $G$, its input comprises three components: a random noise, and the labels of the structural complexity level and the loading position. The labels are embedded into the model using one-hot encoding. Specifically, combining a $100 \times 1$ random vector and the labels, the dimension is expanded to $(100+N_{\rm label})\times1\times1$, where $N_{\rm label}$ is the total dimension of the structural complexity label and the loading position label. For the cantilever beam example, it is $6+101=107$, respectively. Then this vector will be filtered with transposed convolution, denoted as CONVT. For $G$, rectified linear unit (ReLU) is used as the activation function. Batch normalization is implemented for the stabilization of the training process. The output of the generator is a $200 \times 200$ pixel image. 

Both real and generated images undergo the embedding of structural complexity and loading position labels in the same manner. Subsequently, they are input into the discriminator $D$, which, through a series of convolution operations, outputs the Wasserstein distance. Specifically, the input layer of the discriminator is an image sample (either real or generated) with a dimension of $200 \times 200$ and labels of structure complexity and loading position. The following layers of the discriminator are a series of convolutional layers (CONV) paired with batch normalization and leaky rectified linear units (LeakyReLU). 

\subsubsection{Improved loss function}
Training of the WGAN can then be interpreted as a two-player minimax game between the discriminator and the generator. The proposed loss function is expressed as:
\begin{equation}
\begin{aligned}
&\min _G \max _{D} \underbrace{\underset{\tilde{\boldsymbol{x}} \sim \mathbb{P}_g}{\mathbb{E}}[D(\tilde{\boldsymbol{x}}|y_{\rm true})]-\underset{\boldsymbol{x} \sim \mathbb{P}_r}{\mathbb{E}}[D(\boldsymbol{x}|y_{\rm true})]}_{\text {Critic loss of real labels}} \\
& + \underbrace{\lambda \underset{\hat{\boldsymbol{x}} \sim \mathbb{P}_{\hat{\boldsymbol{x}}}}{\mathbb{E}}\left[\left(\left\|\nabla_{\hat{\boldsymbol{x}}} D(\hat{\boldsymbol{x}}|y_{\rm true})\right\|_2-1\right)^2\right] }_{\text {Gradient penalty }} \\ 
& + \underbrace{\underset{{\boldsymbol{x}} \sim \mathbb{P}_r}{\mathbb{E}}[D({\boldsymbol{x}}|y_{\rm false})]}_{\text {Critic loss of fake labels}}
\label{eq12}
\end{aligned}
\end{equation}
where $\mathbb{P}_r$ is the distribution of truly optimized designs, $\mathbb{P}_g$ is the distribution of generated images, $\boldsymbol{x}$ is a real optimized designs, and $\tilde{\boldsymbol{x}}$ is a generated image. $\mathbb{E}$ is the probability measure. $\mathbb{P}_{\hat{\boldsymbol{x}}}$ is the distribution uniformly sampled on a straight line between $\mathbb{P}_r$ and $\mathbb{P}_g$, and $\hat{\boldsymbol{x}} \leftarrow \epsilon \boldsymbol{x}+(1-\epsilon) \tilde{\boldsymbol{x}}$,$\epsilon \sim U(0,1)$. $\lambda$ is a strength factor on the gradient penalty and fixed as $\lambda = 10$ in this paper. $\nabla_{\hat{\boldsymbol{x}}} D(\hat{\boldsymbol{x}}|y_{\rm true})$ is the gradient of the discriminator on the sample $\hat{\boldsymbol{x}}$, and $y_{\rm true}$ is the true information of loading position and complexity level, while $y_{\rm false}$ is the false information. Meanwhile, due to the requirements of conditional Generative Adversarial Networks, it is necessary to introduce conditional probability. 

The above formula mainly consists of three parts. The first part comes from the original WGAN \citep{arjovsky2017wasserstein}, which proposes weight clipping to ensure Lipschitz constraint; however, it leads to optimization difficulties and causes more instability issues in training. The main reason is that clipping is highly sensitive to the values chosen to clip and will likely yield convergence issue or training difficulty. 

This issue can be partially addressed by penalizing the gradients, and the so-called WGAN-GP was proposed \citep{gulrajani2017improved}. WGAN-GP proposes an alternative way to enforce the Lipschitz constraint. It enforces a soft version of the constraint with a penalty on the gradient norm for random samples $\hat{\boldsymbol{x}} \sim \mathbb{P}_{\hat{\boldsymbol{x}}}$, which constitutes the second part of the cost function. 

In the third part, we introduce the critic loss part of fake labels. During the training process, it was found that simply introducing the loss for samples with true loading position could not effectively control the loading position of the generated designs. Therefore, the WGAN loss was further modified by adding a penalty term for real samples with incorrect loading positions. Such samples are generated by a truly optimized design with a randomly selected incorrect loading position. 
\begin{figure*}
	\centering
		\includegraphics[scale=.3]{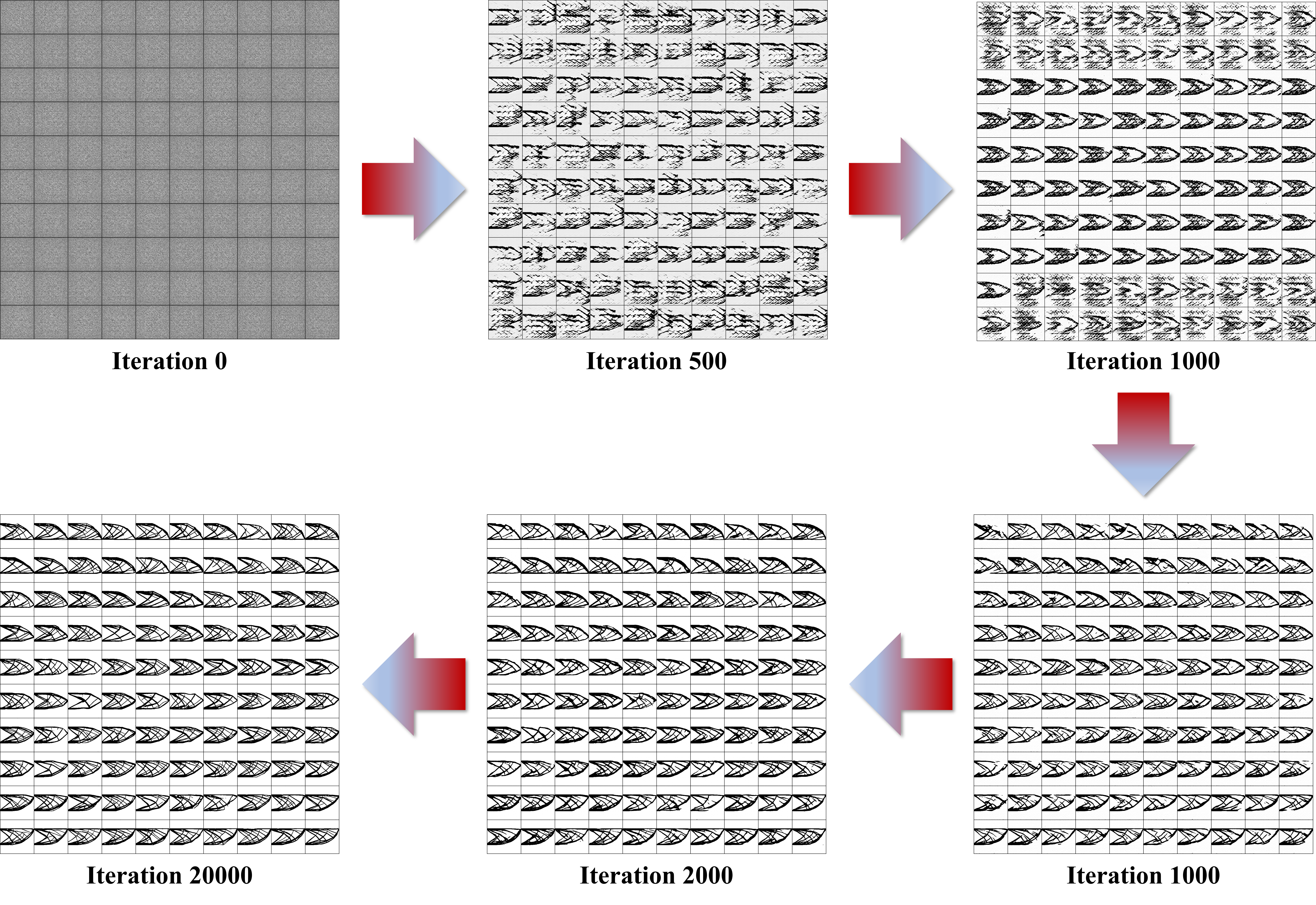}
	\caption{Generated samples from loading positions 1 to 100 by the generator with fixed noises at different training iterations.}
	\label{FIG:8}
\end{figure*}

\subsection{Training process of the proposed WGAN}
Training of the WGAN is carried out with the ADAM algorithm based on the PyTorch library. During the training process, the batch size was 128 and it took 21 hours and 32 minutes to complete the training on a NVIDIA GeForce RTX 3090 GPU. 

Fig. \ref{FIG:8} shows some representative generated images of the generator during the training process of the proposed WGAN model. By fixing 100 random noise vectors and loading position labels ranging from 1 to 100 (i.e., loading from the bottom right corner to the top right corner), images composed of the corresponding 100 generated structures are compared as Fig. \ref{FIG:8}. The evolution process of generated samples can be divided into the following stages:
\begin{itemize}
 \item At iteration 0, the generator only outputs chaotic noise pictures.
 \item At iteration 500, blurred structures begin to emerge.
 \item At iteration 1000, load transmission paths of generated structures become clear.
 \item At iteration 1500, reasonable clear structures matching with the loading positions are observed.
 \item From iterations 2000 to 20000, optimized designs with gradually clearer structural details and  boundary are generated.
\end{itemize}
This indicates that the improved WGAN loss effectively controls the embedded information during the training process. In fact, in early attempts, without introducing the third loss term, the loading position label could not effectively control the generated designs even with more training iterations. 

\section{Numerical examples}
To validate the effectiveness of the proposed WGAN model, we conducted two numerical examples: a cantilever beam example and an L-shaped beam example with a non-design region. The dataset was constructed individually, and separate training was conducted for each dataset. Numerical examples demonstrated the proposed design paradigm is able to generate diverse and almost "truly" optimized structures with controlled loading position and complexity levels for various problem settings. For illustration purpose, all parameters involved are dimensionless, with the Young's modulus and Poisson's ratio setting as 1 and 0.3, respectively. The volume fraction of both examples is fixed as 0.35.
\subsection{The cantilever beam example}
As stated in subsection 2.4, the rectangular design domain of cantilever beam example depicted in Fig. \ref{FIG:2}(a) is discretized by $200\times100$ quadrilateral elements. A unit downward concentrated load is applied at the right side of the design domain, with the 101 possible loading positions labeled by 0 to 100, respectively. Finally, 19,015 samples were constructed by topology optimization, with the complexity categorized into 6 levels as Fig. \ref{FIG:4}. By extending the pixel dimensions of the optimized design as $200 \times 200$, the WGAN model is trained by 20,000 iterations.

\begin{figure*}
	\centering
		\includegraphics[scale=.2]{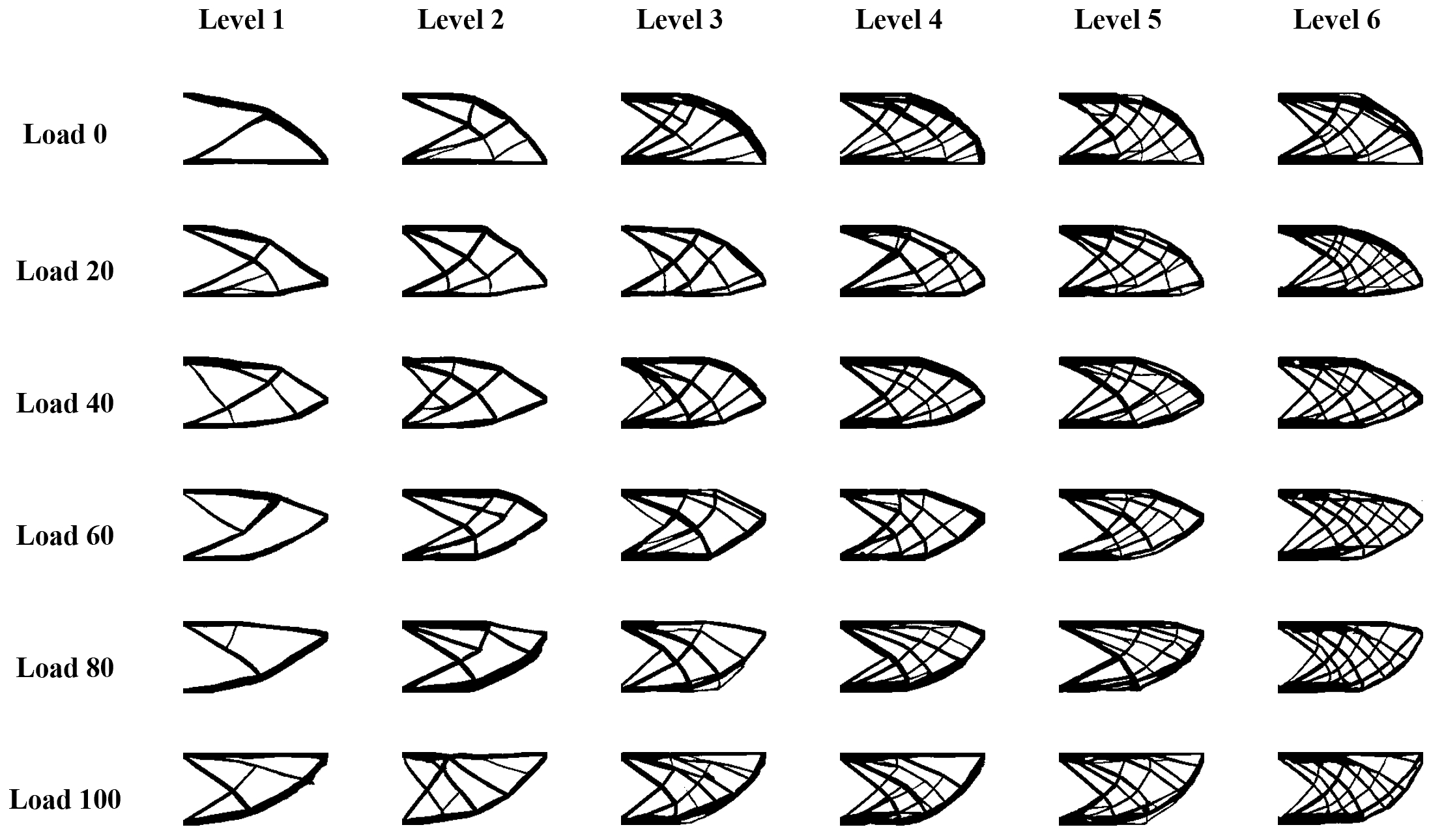}
	\caption{Generated optimized cantilever beams labeled by different complexity levels and loading positions.}
	\label{FIG:9}
\end{figure*}

This well-trained WGAN model could produce highly qualified optimized designs with respect to the loading position and structural complexity labels. Fig. \ref{FIG:9} illustrates generated structures of varying complexity levels acquired at different loading positions. It can be found that, as the label of loading position increases from 0 to 100, the right corner of generated designs evolves from the bottom right to the top right; as the complexity level increases from 1 to 6, crisp optimized structures are generated with more structural details. 

\begin{figure*}
	\centering
		\includegraphics[scale=.2]{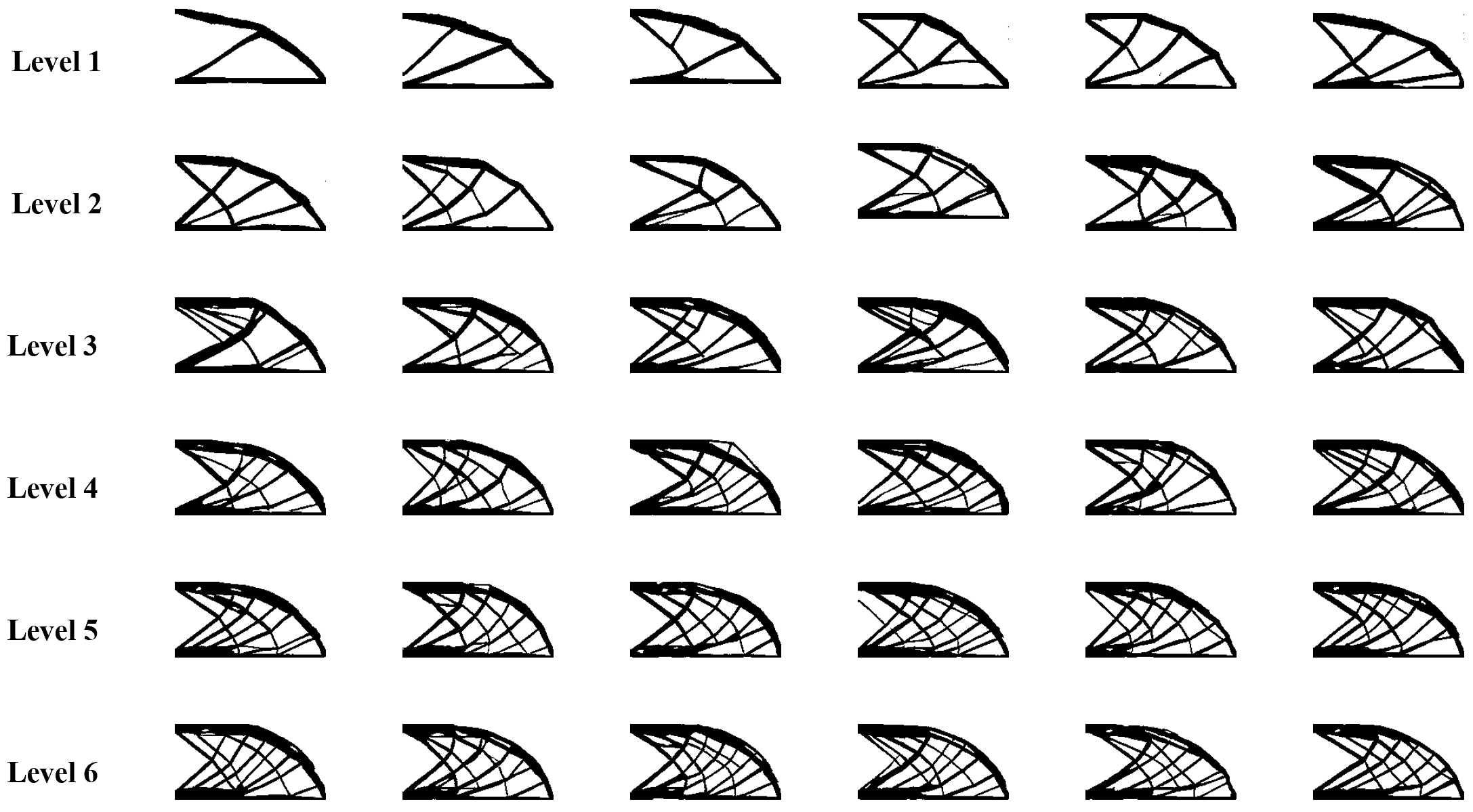}
	\caption{Generated diverse optimized cantilever beams labeled by different complexity levels under the loading position 0.}
	\label{FIG:10}
\end{figure*}
Furthermore, for the loading position label of 0, Fig. \ref{FIG:10} presents a set of generated designs with different complexity levels. Again, besides showing a clear load transmission path and crisp boundary, all those generated samples have very good agreement with the labels of loading position and complexity level. Compared with the corresponding training samples presented in Fig. \ref{FIG:5}, some new design results not appeared in the training set are produced. This illustrates the generalization ability of the proposed WGAN model.

It is worth noting that, although the generated designs are pixel images, their quality is very high and this is not easy to be achieved by most existing generative models for structural topology optimization. In order to quantitatively evaluate the sharpness of the generated samples, the so-called Measure of Non-Discreteness ($M_{nd}$) \citep{sigmund2007morphology} are adopted here:\begin{equation}
M_{nd}=\frac{\sum_{e=1}^n 4 \tilde{\rho}_e\left(1-\tilde{\rho}_e\right)}{n} \times 100 \%
\label{eq13}
\end{equation}
If a design is entirely discrete (purely black-and-white), the Measure of Non-Discreteness ($M_{nd}$) is 0\%. Conversely, if the elemental densities is homogeneous 0.5, $M_{nd}$ is 100\%. 
\begin{figure*}
	\centering
		\includegraphics[scale=.475]{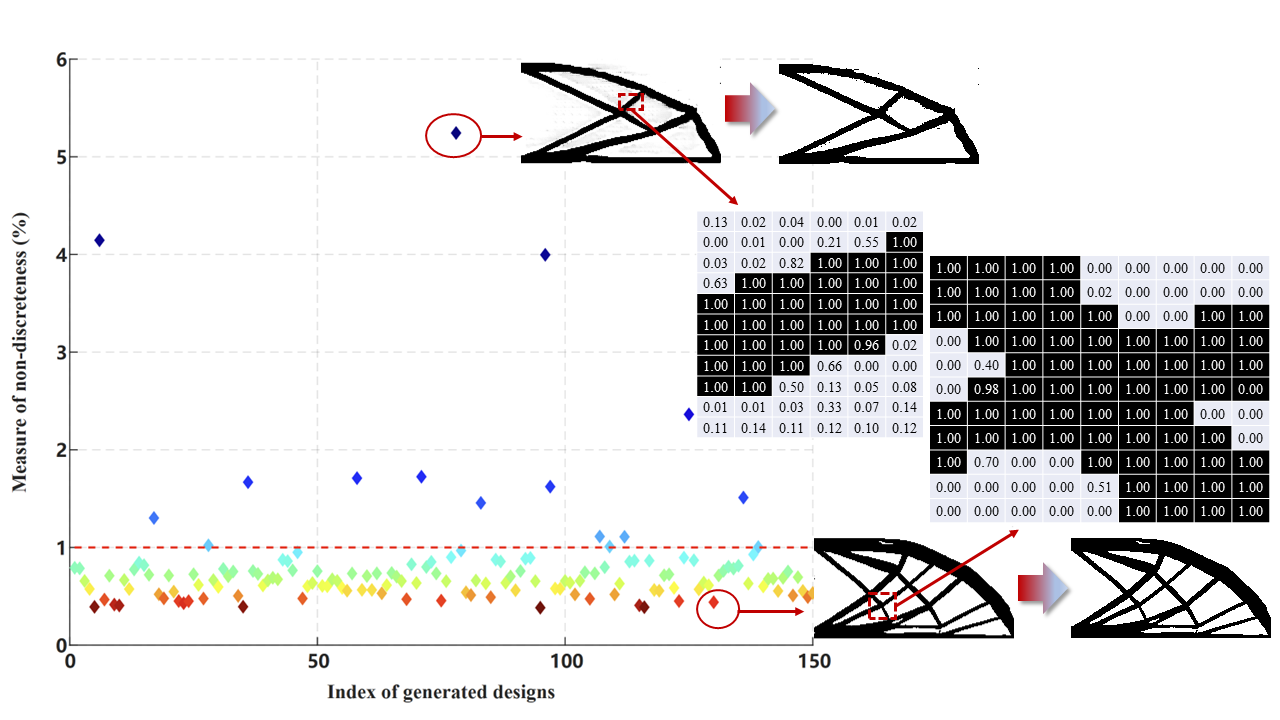}
	\caption{Discreteness analysis of generated structures at loading position 0.}
	\label{FIG:11}
\end{figure*}

\begin{figure}
	\centering
		\includegraphics[scale=0.7]{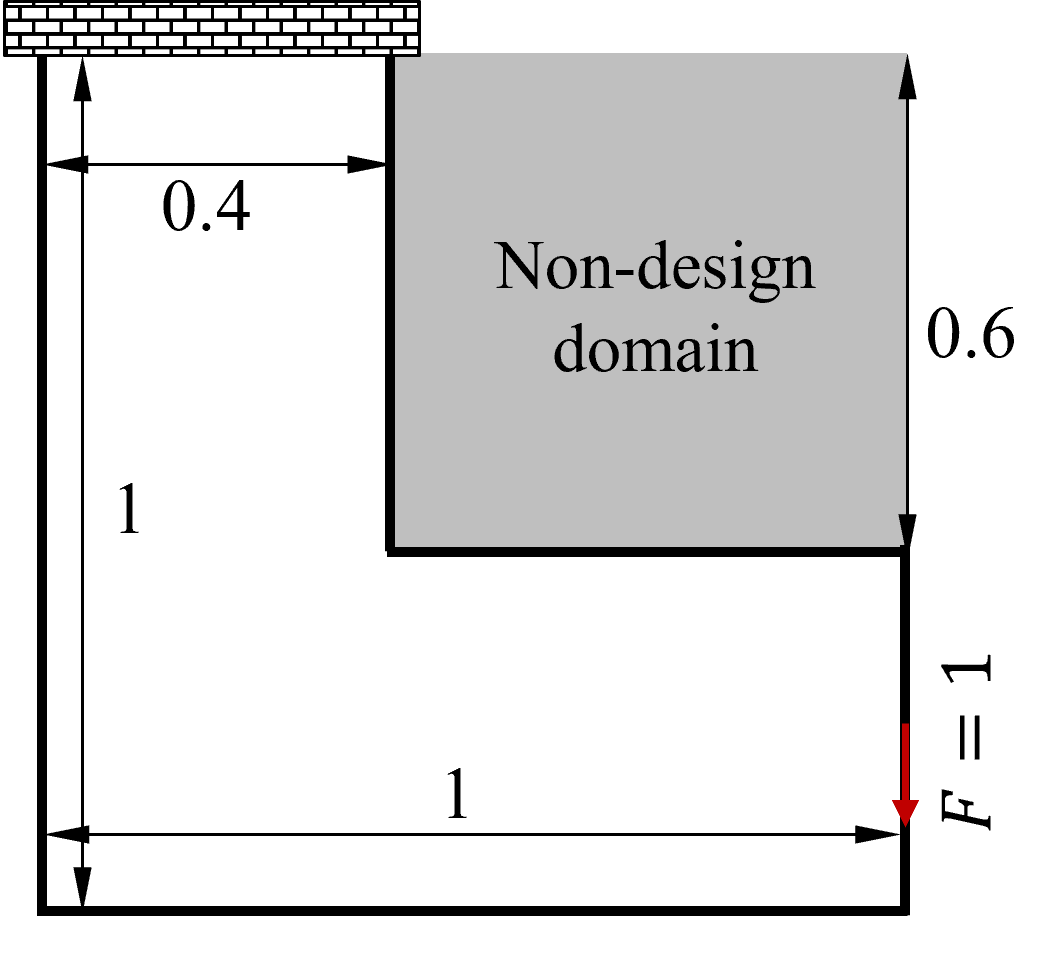}
	\caption{Problem setting of the L-shaped beam example.}
	\label{FIG:12}
\end{figure}

\begin{figure}
	\centering
		\includegraphics[scale=0.5]{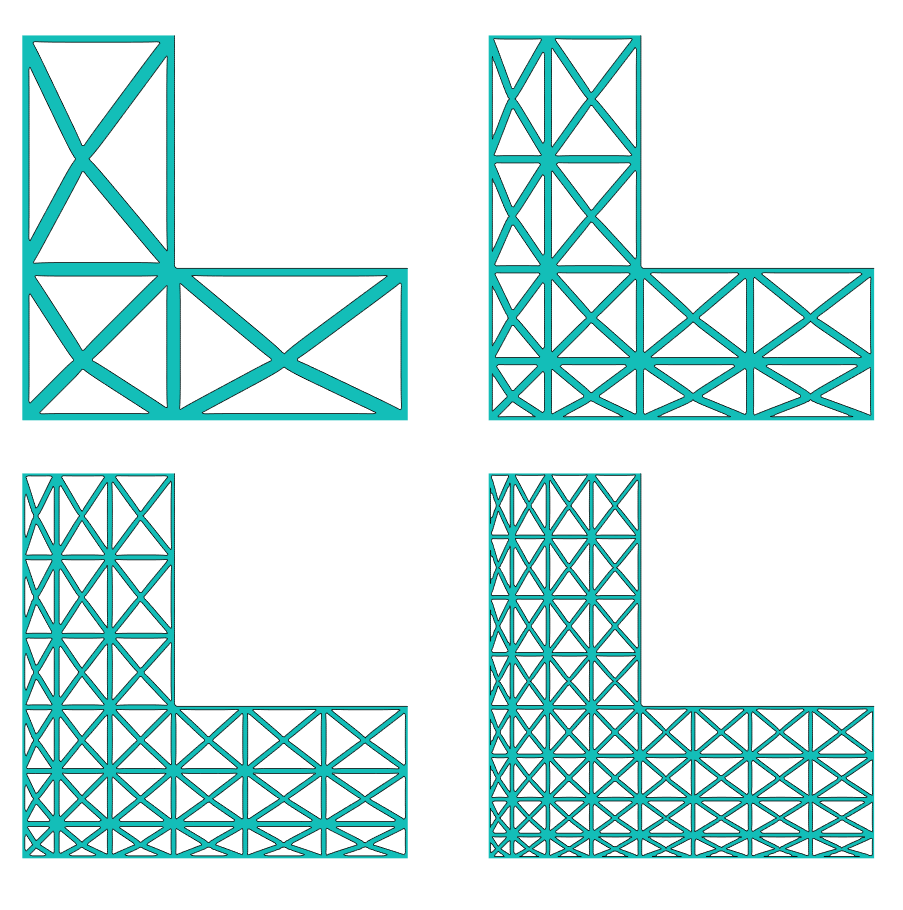}
	\caption{Some representative initial designs of the L-shaped beam example.}
	\label{FIG:13}
\end{figure}

By randomly generating 150 outputs with loading position 0 (loading at the bottom right corner), their Measure of Non-Discreteness was calculated using Eq. (\ref{eq13}). The scatter diagram in Fig. \ref{FIG:11} illustrates that most generated designs have an $M_{nd}$ value below 1\%, indicating a small number of gray elements. The figure also showcases two typical structures, one with the largest value $M_{nd}$ value (poor sharpness) and another with a low $M_{nd}$ value. For the most poor sample with $M_{nd}$ exceeding 5\%, gray elements are observed in the "void" region, nevertheless, the load transmission path is quite clear. Therefore, discrete optimized structures can be obtained by setting a truncation threshold, e.g., 0.85.

This highlights the significance of utilizing the modified MMC method to generate dataset. Thanks to the sharpness and a clear load transmission path in the training samples, the generated designs exhibit the same advantages. These characteristics provide engineers more intuitive references and simplify the post-processing. 

\subsection{The L-shaped beam example  with a non-design region}
The dimensions and boundary conditions of the L-shaped beam example are depicted in Figure \ref{FIG:12}. The unit square domain is divided into $200 \times 200$ quadrilateral elements with the non-design domain fixed as $120 \times 120$ void elements. The totally 81 possible uniform loading positions are labeled by 0 to 80 from the bottom right corner of the square to the bottom right corner of the non-design region. From the initial designs illustrated by Fig. \ref{FIG:13}, combining the strategies in subsection 2.2, totally 8,672 optimized designs were constructed, with the complexity categorized into 3 levels (genus numbers: 0-6 for level 1, 7-12 for level 2, 13 and more for level 3). The WGAN model in Fig. \ref{FIG:7} is trained by 20,000 iterations as well.
\begin{figure*}
	\centering
		\includegraphics[width = 0.9\textwidth]{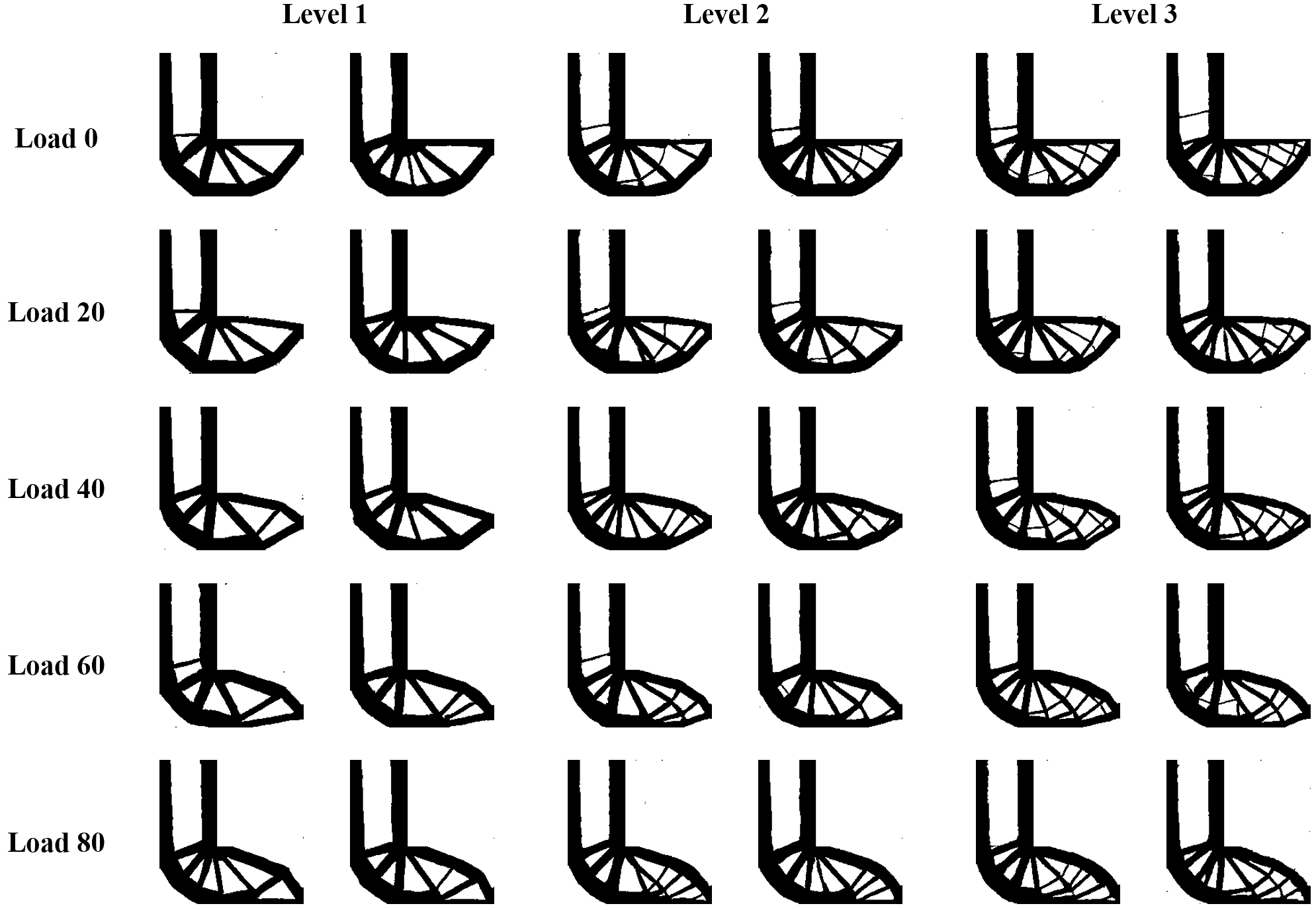}
	\caption{Generated optimized L-shaped beams labeled by different complexity levels and loading positions.}
	\label{FIG:14}
\end{figure*}

\begin{figure*}
	\centering
		\includegraphics[width = 0.9\textwidth]{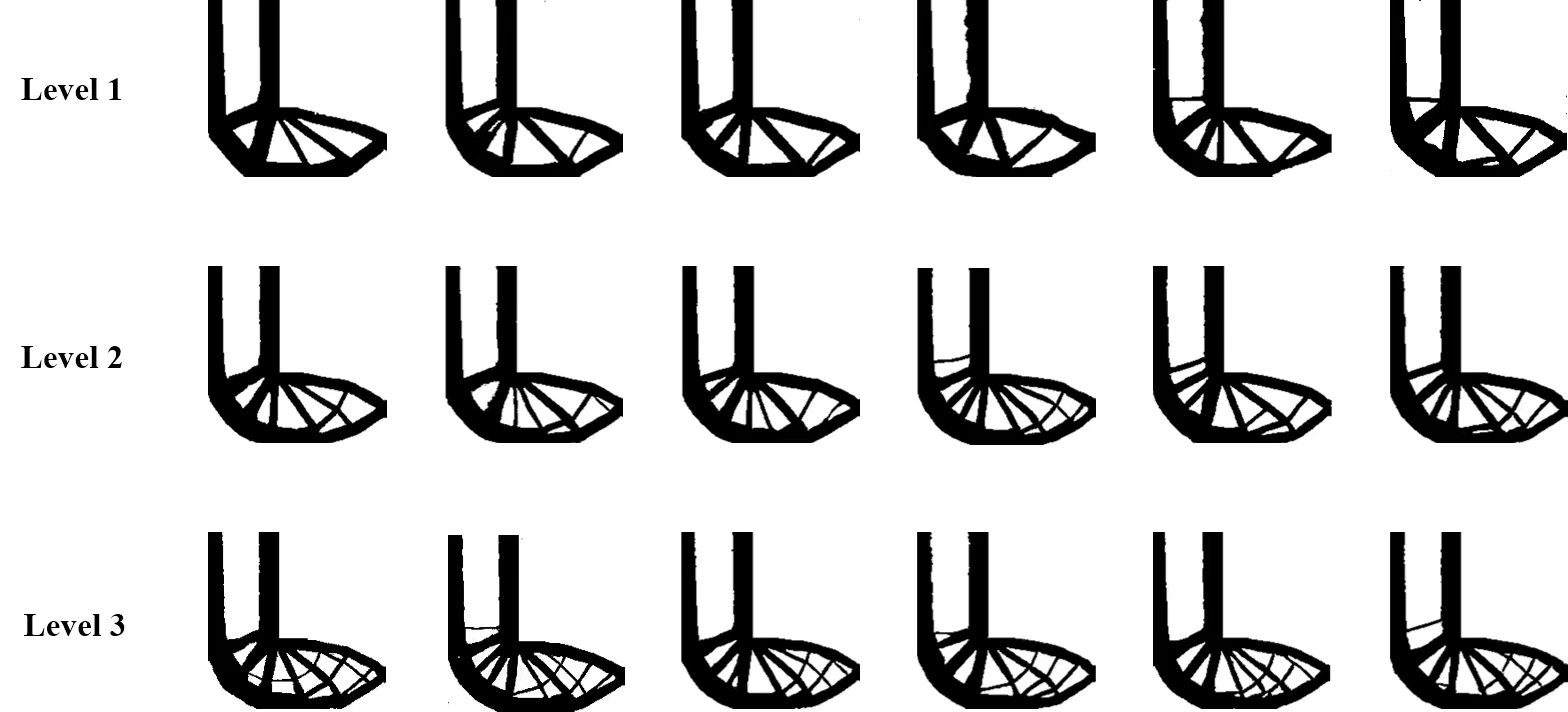}
	\caption{Generated diverse optimized L-shaped beams labeled by different complexity levels at loading position 40.}
	\label{FIG:15}
\end{figure*}

To demonstrate the reliability of generated designs, Fig. \ref{FIG:14} presents generated structures corresponding to different complexity levels and loading positions. By double checking the genus and the right-hand-side edge of generated configurations, the real complexity level matches exactly while the loading point is always surrounded by solid elements and falls in the load transmission path. In addition, in Fig. \ref{FIG:15}, by fixing the loading position label as 40, i.e., at the center of the right edge, diverse structures of varying complexity levels are generated. Notably, as the complexity level label increases, optimized L-shaped beams with more structural details are generated, and the non-design domain is also well preserved as void. 

To quantitatively evaluate the structural performance of the optimized designs, Table \ref{tbl4} presents the reanalysis results of six pairs of reference designs in dataset and the corresponding generated designs similar to them. Interestingly, besides generating similar configurations usually stated in literature about generative models, the genus (number of holes), the volume fraction, and the objective function of generated samples are very close to the corresponding reference designs, with relative differences about 1\% or even smaller. 

This example demonstrates that, on the one hand, the proposed WGAN model is able to produce diverse optimized designs with controllable structural complexity and loading position in real-time; on the other hand, thanks to the high-quality dataset and improved loss function, almost "truly" optimized designs with a clear load transmission path and crisp boundary can be produced. These advantages would make the proposed generative model favorable for engineers.

\begin{table*}[width=2.05\linewidth,cols=7,pos=h]
\caption{Quantitatively evaluation of the generated designs with their reference.}\label{tbl4}
\begin{tabular*}{\tblwidth}{@{} CCCCCC@{} }
\toprule
Reference & Output & Reference & Output& Reference & Output
\\
\midrule
\\
 \begin{minipage}[b]{0.25\columnwidth}
    \centering
    \raisebox{-.5\height}{\includegraphics[width=\linewidth]{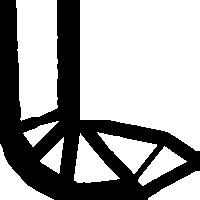}}
\end{minipage} & \begin{minipage}[b]{0.25\columnwidth}
    \centering
    \raisebox{-.5\height}{\includegraphics[width=\linewidth]{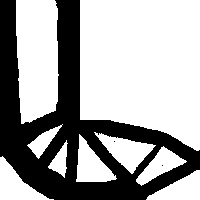}}
\end{minipage} & \begin{minipage}[b]{0.25\columnwidth}
    \centering
    \raisebox{-.5\height}{\includegraphics[width=\linewidth]{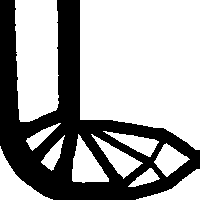}}
\end{minipage} & \begin{minipage}[b]{0.25\columnwidth}
    \centering
    \raisebox{-.5\height}{\includegraphics[width=\linewidth]{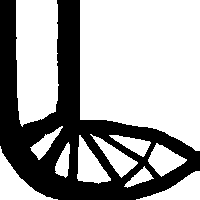}}
\end{minipage}  & \begin{minipage}[b]{0.25\columnwidth}
    \centering
    \raisebox{-.5\height}{\includegraphics[width=\linewidth]{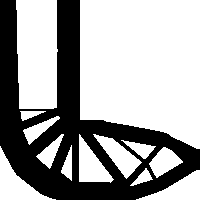}}
\end{minipage} & \begin{minipage}[b]{0.25\columnwidth}
    \centering
    \raisebox{-.5\height}{\includegraphics[width=\linewidth]{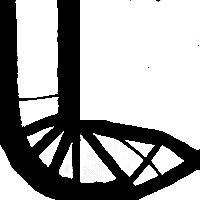}}
\end{minipage} 

\\\\

\makecell[c]{ Hole: 5 \\  Vol: 0.356 \\ Obj: 344.3 } &\makecell[c]{ Hole: 5 \\Vol: 0.354\\ Obj: 346.0} &\makecell[c]{ Hole: 7\\ Vol: 0.351\\ Obj: 339.6}&\makecell[c]{Hole: 7 \\Vol: 0.351\\ Obj: 343.5} & \makecell[c]{ Hole: 9 \\  Vol: 0.359 \\ Obj: 339.9 } &\makecell[c]{ Hole: 9 \\Vol: 0.356\\ Obj: 340.9} 

\\\\

 \begin{minipage}[b]{0.25\columnwidth}
    \centering
    \raisebox{-.5\height}{\includegraphics[width=\linewidth]{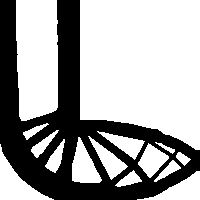}}
\end{minipage} & \begin{minipage}[b]{0.25\columnwidth}
    \centering
    \raisebox{-.5\height}{\includegraphics[width=\linewidth]{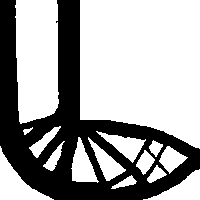}}
\end{minipage} & \begin{minipage}[b]{0.25\columnwidth}
    \centering
    \raisebox{-.5\height}{\includegraphics[width=\linewidth]{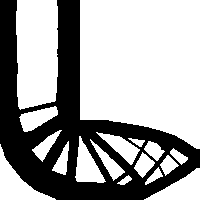}}
\end{minipage} & \begin{minipage}[b]{0.25\columnwidth}
    \centering
    \raisebox{-.5\height}{\includegraphics[width=\linewidth]{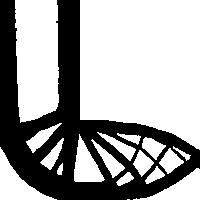}}
\end{minipage}  & \begin{minipage}[b]{0.25\columnwidth}
    \centering
    \raisebox{-.5\height}{\includegraphics[width=\linewidth]{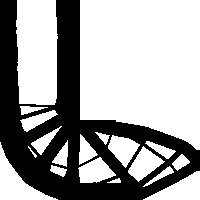}}
\end{minipage} & \begin{minipage}[b]{0.25\columnwidth}
    \centering
    \raisebox{-.5\height}{\includegraphics[width=\linewidth]{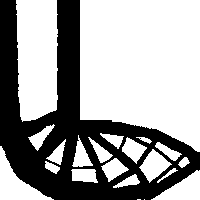}}
\end{minipage} 
\\\\
\makecell[c]{ Hole: 11\\ Vol: 0.355\\ Obj: 338.2}&\makecell[c]{Hole: 11 \\Vol: 0.356\\ Obj: 338.1} &\makecell[c]{ Hole: 13 \\  Vol: 0.358 \\ Obj: 335.8 } &\makecell[c]{ Hole: 13 \\Vol: 0.355 \\ Obj: 334.2} &\makecell[c]{ Hole: 15\\ Vol: 0.362\\ Obj: 334.3}&\makecell[c]{Hole: 15 \\Vol: 0.365\\ Obj: 334.3}

\\\bottomrule
\end{tabular*}
\end{table*}

\section[Theorem and ...]{Concluding remarks}
The present study proposes a real-time framework for generating diverse optimized designs with controllable complexity based on WGAN. The trained model immediately generates a set of "truly" optimized structures by inputting random noises along with loading positions and complexity level. Main contributions of this work are summarized as:

First, a modified MMC method that uses movable nodes and components to represent structures is adopted, resulting in optimized structures that exhibit crisp boundary and a clear load transmission path. With three strategies on controlling the layout of components and nodes, a high quality dataset is produced and shared on Github.

Second, complexity level based on the topological invariant of optimized structures is proposed to measure the complexity and set as a label of the WGAN model to control the complexity of optimized designs. Additionally, a novel error term is introduced to successfully match the loading position label with the output images.

Finally, the proposed framework is validated for optimal design of a cantilever beam example and a L-shaped beam example with a non-design region. It is important to note that the framework not only reproduces almost "truly" optimized structures in the dataset but also generates novel configurations that meet the desired demands. 

The present design framework can be extended for three-dimensional topology optimization and further incorporated with subsequent design and application process to give full play of the intelligence design. 

\section*{Declaration of Competing Interest}
The authors declare that they have no known competing financial interests or personal relationships that could have appeared to influence the work reported in this paper.

\section*{Acknowledgements}
The National Natural Science Foundation (11821202, 12372122 and 12202092), the Science Technology Plan of Liaoning Province (2023JH2/101600044), Liaoning Revitalization Talents Program (XLYC2001003), and 111 Project (B14013) are gratefully acknowledged.

%\section*{Data availability}
%The training dataset is shared in Github. Other data will be made available from the corresponding authors upon reasonable request.
%\bibliographystyle{cas-model2-names}
\bibliographystyle{unsrtnat}

\bibliography{cas-refs}

\end{document}